%% file: main.tex
\documentclass{nature}

\usepackage{ulem}
\usepackage{xcolor}

\title{A biology-driven deep generative model for cell-type annotation in cytometry}

\author{Quentin Blampey$^{1,2}$, Nadège Bercovici$^{2,3}$, Charles-Antoine Dutertre$^{2,4}$, Isabelle Pic$^2$, Fabrice André$^2$, Joana Mourato Ribeiro$^2$ \& Paul-Henry Cournède$^1$}

\usepackage{amsmath}
\renewcommand{\vec}[1]{\boldsymbol{#1}}
\usepackage{amssymb}
\usepackage{graphicx}
\usepackage[skip=2pt,font=scriptsize]{caption}

\DeclareMathOperator*{\argmax}{arg\,max}

\usepackage{booktabs, 
            makecell}

\usepackage[colorlinks = true,
            linkcolor = black,
            urlcolor  = black,
            citecolor = black,
            anchorcolor = black]{hyperref}
\usepackage{dsfont}

\begin{document}

\maketitle

\begin{affiliations}
 \item Paris-Saclay University, CentraleSupélec, Laboratory of Mathematics and Computer Science (MICS), Gif-sur-Yvette, 91190 France
 \item Paris-Saclay University, Gustave Roussy, INSERM U981, PRISM Center, Villejuif, France
 \item Université Paris Cité, Institut Cochin, INSERM, CNRS, Paris, 75014 France
 \item Institut National de la Santé Et de la Recherche Médicale (INSERM) U1015, Equipe Labellisée—Ligue Nationale contre le Cancer, Villejuif, France.
\end{affiliations}

\begin{abstract}
Cytometry enables precise single-cell phenotyping within heterogeneous populations. These cell types are traditionally annotated via manual gating, but this method suffers from a lack of reproducibility and sensitivity to batch effect. Also, the most recent cytometers — spectral flow or mass cytometers — create rich and high-dimensional data whose analysis via manual gating becomes challenging and time-consuming.
To tackle these limitations, we introduce Scyan
\footnote{\url{https://github.com/MICS-Lab/scyan}}, a \underline{S}ingle-cell \underline{Cy}tometry \underline{A}nnotation \underline{N}etwork that automatically annotates cell types using only prior expert knowledge about the cytometry panel. For this, it uses a normalizing flow — a type of deep generative model — that maps protein expressions into a biologically relevant latent space.
We demonstrate that Scyan significantly outperforms the related state-of-the-art models on multiple public datasets while being faster and interpretable. In addition, Scyan overcomes several complementary tasks, such as batch-effect correction, debarcoding, and population discovery. Overall, this model accelerates and eases cell population characterisation, quantification, and discovery in cytometry.
\end{abstract}

\input{1_intro}
\input{2_results}

\input{3_discussion}
\input{4_methods}

\section{Acknowledgement}
This work is supported by Prism – National Precision Medicine Center in Oncology funded by the France 2030 programme and the French National Research Agency (ANR) under grant number ANR-18-IBHU-0002.

\section{Code availability}
The code developed in this article is available as an open-source Python package, accessible on Github at \url{https://github.com/MICS-Lab/scyan}. The code used to run and compare the other algorithms is also available, at \url{https://github.com/quentinblampey/cytometry_benchmark}.

\section{Data availability}
The four datasets and knowledge tables considered in this paper are public and accessible at \url{https://github.com/MICS-Lab/scyan_data}

\input{6_supp}

\end{document}

%% file: 1_intro.tex
\section{Introduction}

The simultaneous detection of several cellular proteins by spectral and mass cytometry opens up an unprecedented way to detect, quantify, and monitor the function of highly specific cell populations from complex biological samples\cite{cyto2}. These rich analyses are made possible with the usage of large panels of markers, typically more than 30 or 40 markers, which considerably increases the information contained in the data\cite{cyto1}. They provide key insights to better understand specific diseases, immune cell functions, or monitor the response to therapies\cite{flow_overview}.
To obtain such results, population annotation must be performed to provide each cell with a biologically meaningful cell type. Yet, due to the data's high dimensionality and complexity, manual annotations become challenging and labour intensive\cite{cyto_limitations1}. This process, called gating\cite{gating}, is highly subjective and sensitive to the batch effect, or non-biological data variability\cite{cyto_limitations1}. These drawbacks are amplified as the number of cytometry samples increases, reinforcing the need to develop and use automatic tools in population annotation and data analysis\cite{nat_comment, critical_assessment}.

Many clustering tools\cite{phenograph, leiden, spade} have been developed for automatic data exploration and population discovery. However, a manual analysis of marker expressions is still required to name each cluster with a meaningful cell type. Indeed, clusters do not necessarily correspond to one specific cell type, and it is up to the investigator to decide to which population each cluster corresponds. Clustering tools are also not scaling well on large datasets, and are sensible to batch-effect. An alternative approach to clustering is to use automatic annotation models. The first category of annotation models are supervised or semi-supervised models\cite{deepcytof, lda, cyanno, review}. They rely on prior manual gating of large datasets to train the models. Moreover, these models can only annotate populations with predefined types of cells, they cannot be used to discover new ones. The second category, to which our model belongs, corresponds to unsupervised annotation models that leverage prior biological knowledge about the panel of markers. Although some models have been developed\cite{acdc, mp, review}, they either (i) lack interpretability, (ii) cannot discover new populations, (iii) require the usage of batch-effect correction models before being applied, or (iv) scale poorly to large datasets.
Surprisingly, deep learning has been underused for cytometry annotations, while proving efficient and flexible for many related applications of single-cell biology\cite{scvi, cellassign, saucie}.

In this paper, we introduce a single-cell cytometry annotation network called Scyan that annotates cell types and corrects batch effects concurrently without any label or gating needed.
Scyan is a Bayesian probabilistic model composed of a deep invertible neural network called a normalizing flow\cite{nf1, nf2, nf3}. This flow transforms cell data into a latent space that is used for annotation, does not contain batch effect, and is key for population discovery.

We demonstrate Scyan efficiency, scalability, and interpretability on three public mass cytometry datasets for which manually annotated cell populations are considered ground truth. We compare Scyan classification performance to two knowledge-based approaches\cite{acdc, mp}, one clustering method\cite{phenograph, acdc}, and two supervised models\cite{lda, cyanno}. Additionally, we compare Scyan batch-effect correction to four state-of-the-art batch correction methods\cite{harmony, cydar, combat, saucie}. We also show that our model can be used for population discovery, as well as for the general task of debarcoding. Overall, these properties make Scyan an end-to-end analysis framework for mass/spectral/flow cytometry.

%% file: 2_results.tex
\section{Results}

\section*{Scyan model architecture}

\begin{figure*}[hbt!]
\centering
\includegraphics[width=\linewidth]{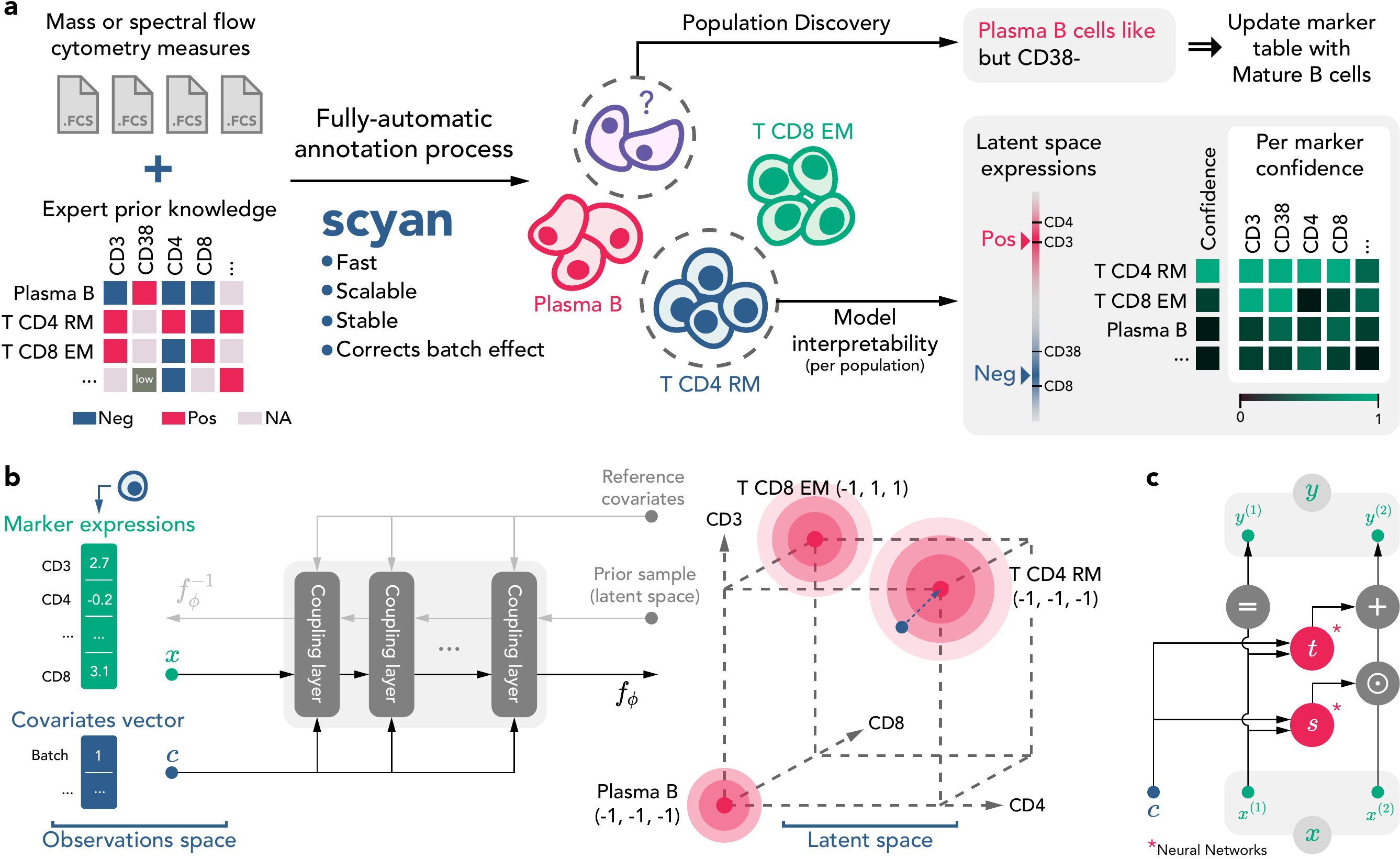}
\caption{\label{fig:1}\textbf{Overview of Scyan usage and architecture.} \textbf{a}, Illustration of Scyan typical use case. It requires (i) one or multiple cytometry acquisitions and (ii) a table that details which population is expected to express which markers. Then, Scyan annotates cells in a fast and unsupervised (or fully-automatic) manner while removing batch effect (if any). After training, we provide interpretability tools to understand Scyan annotations and discover new populations that can eventually be added to the table afterwards. \textbf{b}, Illustration of Scyan architecture. One cell is represented by its marker expressions vector and eventual covariates; then a deep generative model (in particular, a normalizing flow) named $f_{\Vec{\phi}}$ maps these two vectors into a latent space. The latent space has the same shape as the original space, and is built using the biological knowledge table. Once a cell is mapped into the latent space, annotation can be made by choosing the highest probable population, whose distribution is Gaussian-like and on a hypercube vertex. \textbf{c}, One coupling layer, the elementary unit that composes the transformation $f_{\Vec{\phi}}$, contains two multi-layer perceptrons ($s$ and $t$) and uses cell covariates such as the batch information.}
\end{figure*}

Scyan is composed of two core components: (i) $f_{\vec{\phi}}$, a neural network called normalizing flow, and (ii) a latent space on which a target distribution $\vec{U}$ is defined (\autoref{fig:1}b). This target distribution is a mixture of distributions — one per population — built using prior biological knowledge about the cell types. This knowledge is provided as a table: for all populations, each expected marker expression is given or left unknown (more details in supplementary \autoref{table_details}). This table is then used to define mathematically the target distribution $\vec{U}$. Also, the latent space (on which $\vec{U}$ is defined) has the same dimension as the original space; therefore, each marker has its corresponding latent expression.\\

The purpose of the normalizing flow is to learn an invertible mapping between the actual marker expression distribution and the target $\vec{U}$. By mapping marker expressions to a biologically-defined latent space, we force the transformation to provide latent expressions on a scale that is shared for every marker, going from negative (-1) to positive (+1). These latent marker expressions are meant to be free of batch effect or any non-biological factor.  
By the design of $f_{\vec{\phi}}$ and of the objective function, the normalizing flow is not allowed to make huge space distortions, which helps preserve the biology.

Annotations are performed on the latent space after learning the model parameters $\vec{\phi}$. We annotate a cell by choosing the population distribution whose likelihood is the highest for the cell latent representation. If a cell latent representation does not correspond to any component of the mixture, then the cell remains unlabelled, but population discovery can be run afterwards to annotate it eventually.

\section*{Scyan provides a better and faster annotation than unsupervised methods}

\subparagraph{Classification metrics comparison}

We compare Scyan to the related works on three public mass cytometry datasets. One is from patients with acute myeloid leukemia\cite{phenograph} (AML, N = 104 184 cells), one from bone marrow mononuclear cells\cite{bmmc} (BMMC, N = 61 725 cells), and the last one from peripheral blood mononuclear cells (PBMCs) samples of peanut-allergic individuals (POISED, N = 4 178 320 cells). The latter is the largest of the three, and contains 30 samples, divided among 7 batches, and under two different conditions (peanut stimulated or unstimulated). More descriptions can be found in supplemental \autoref{supp_dataset}. Manual gating has been performed in previous studies\cite{phenograph, bmmc, poised}, providing ground truth labels to evaluate annotation models. Note that the unsupervised models listed below do not use these labels during training.

\begin{figure*}[hbt!]
\centering
\includegraphics[width=\linewidth]{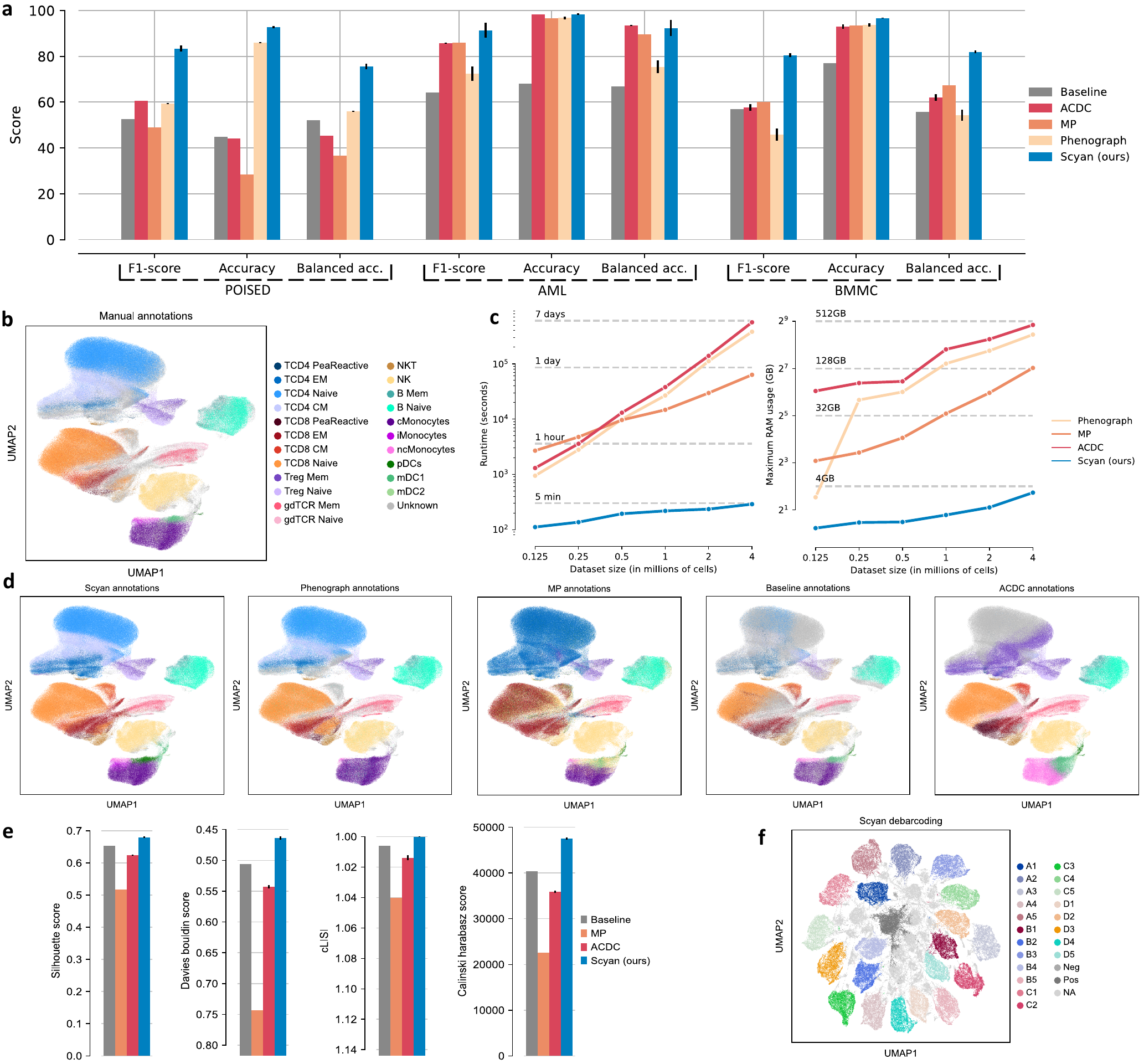}
\caption{\label{fig:2}\textbf{Comparison to state-of-the-art unsupervised methods.} \textbf{a}, Performance comparison of Scyan and four other unsupervised methods on three datasets (POISED, AML, BMMC) using three metrics for each. \textbf{b}, UMAP representing the manually annotated populations on the POISED dataset. \textbf{c}, Models runtime comparison (left) and RAM usage comparison (right) over multiple dataset sizes. \textbf{d}, UMAPs representations of the annotations of all five models on the POISED dataset. \textbf{e}, Unsupervised metrics for the debarcoding task. \textbf{f}, UMAP representing Scyan debarcoding. Cells that did not correspond to any desired barcode were left unclassified (NA).}
\end{figure*}

We compared Scyan to three other knowledge-based models: ACDC\cite{acdc}, a baseline model (defined by the authors of ACDC), and MP\cite{mp}. Also, we compared our model to Phenograph\cite{phenograph}, a clustering model. Note that the Phenograph does not predict labels itself, thus each cluster has to be assigned to a biological cell-type. This is typically done by human experts, but, for more objectivity, the clusters were named using known labels. Using these labels thus replaces the assignment of clusters to biological cell types by human experts and provides a way to compare Phenograph to the other approaches by making the assumption that a human expert would correctly annotate the clusters. We evaluated the models using accuracy, macro averaged F1-score, and balanced accuracy. The results are detailed in \autoref{fig:2}a.

The tests show that Scyan outperforms the other models. In particular, Scyan is about 20 points higher than the other models on POISED and BMMC for the F1-score and the balanced accuracy, which is explained by the capacity of Scyan to better detect small populations (\autoref{fig:2}a). On the BMMC dataset, ten sub-populations represent less than one percent of the total number of cells, making these populations more difficult to detect and label. Yet, small population annotations can still be essential, and thus so is Scyan's capacity to detect them. Also, the gap between Scyan and the other models is more stringent for POISED, showing our model's ability to better annotate large and complex datasets with batch effect.

\subparagraph{Computational speed, scalability, and memory usage}

To demonstrate the scalability of Scyan on large datasets, we compare the execution times and the random access memory (RAM) usage of the different algorithms over multiple dataset sizes (\autoref{fig:2}c). The different sizes were obtained by sub-sampling the POISED dataset, for various sample sizes from 125,000 to 4 million cells. All experiments were run using the same hardware; in our case, CPUs only (i.e. no GPU acceleration, even though Scyan can use GPUs). On $N =$ 4 million cells, Scyan runs in five minutes, while ACDC/MP/Phenograph need between one day and seven days. Scyan scales well to large datasets, as shown by the low slope on \autoref{fig:2}c. This low slope is explained by the fact that Scyan learns a transformation over a space of size $M$ (the number of markers) instead of directly learning a specific annotation for each cell (the actual annotations happen after training). Thus, while the number of cells increases, the spatial complexity does not, and the model remains fast to train. Concerning RAM consumption, Scyan uses less than 4GB of RAM, which means it can be run on any standard laptop. In comparison, ACDC/MP/Phenograph require between 128GB and 512GB of RAM, which is only available on large computer clusters.

\subparagraph{Comparison for barcoding deconvolution}

Barcoding is a method that reduces the batch effect and data variability by allowing the processing of multiple cell samples together, each cell sample being labelled — or barcoded — with a unique combination of antibodies. This protocol requires (i) the dedication of a few markers to make barcodes and (ii) the identification of each cell sample based on its barcode. The latter task, called debarcoding\cite{debarcoding}, can also be expressed as a knowledge-based annotation task. In this situation, we annotate samples instead of populations, and the expert knowledge required for this task simply corresponds to the known barcodes. \autoref{fig:2}e shows that Scyan outperforms ACDC, MP, and the baseline on a public dataset with 20 barcodes and 6 markers\cite{debarcoding}. We added two barcodes corresponding to only negative and only positive markers, with the objective of filtering these cells before further analysis. The UMAP on \autoref{fig:2}f shows a clear separation of the different barcodes, with some small residual clusters (not to be considered) corresponding to non-existing barcodes. The UMAPs corresponding to the debarcoding of the other methods can be found in supplementary \autoref{fig:sup_6}.

\section*{Scyan corrects batch effect}

A batch effect is a phenomenon that induces data variability due to non-biological factors such as the use of a different antibody or slightly different cytometer settings. In practice, these factors may introduce variability that interferes with the analysis and can lead to confusion, over-interpretation, and difficulties in annotating populations. To tackle this issue, Scyan can make some corrections to align the inter-batch distributions. Classically, batch effect correction is performed before annotation, but our method allows for correcting it at the same time as the annotation. Taking into account the batch helps Scyan to annotate the populations better.

\begin{figure*}[hbt!]
\centering
\includegraphics[width=\linewidth]{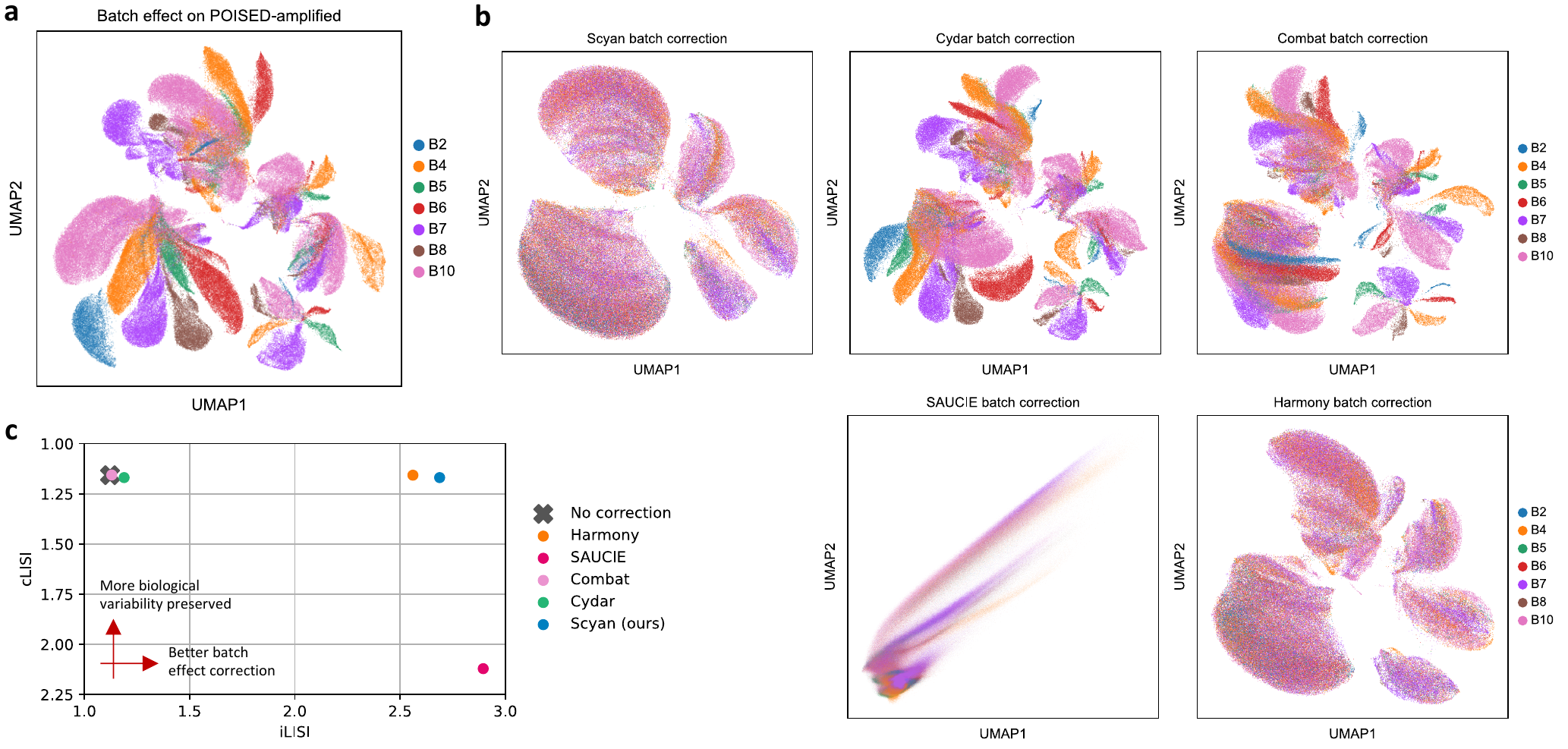}
\caption{\textbf{Batch-effect correction on the POISED dataset with batch-effect amplification.} \label{fig:3}\textbf{a}, UMAP showing the 7 different batches (before batch effect correction). The batch effect is visible since different batches form separated clusters. \textbf{b}, Batch-effect correction of Scyan, Cydar, Combat, SAUCIE, and Harmony. A good batch effect correction can be observed by a superposition of all batch distributions. \textbf{c} Batch-effect correction metrics. A low cLISI (in the top figure) denotes good cell-type variability preservation, while a high iLISI (in the right figure) denotes better batch mixing.}
\end{figure*}

We benchmarked our model ability to correct batch effect to four models: Cydar\cite{cydar}, Combat\cite{combat}, SAUCIE\cite{saucie}, and Harmony\cite{harmony}. We used the POISED dataset on which we had 7 biological batches, and we amplified the batch effect to complex the batch correction (see methods \autoref{bea}). In \autoref{fig:3}c, we provide two metrics: the cell-type LISI (cLISI) that measures if the biological variability is kept, while the integration LISI (iLISI) measures how well the batches overlap, i.e., if the batch-effect was corrected. \autoref{fig:3}b,c show that Cydar's and Combat's corrections are very limited, even though they keep the biological variability. SAUCIE provides the best iLISI, so it mixes well the different batches, but it also removes most of the biological variability (high cLISI). On the opposite, Scyan and Harmony successfully remove the batch effect while preserving the biological variability.

Formally, the batch effect is corrected by integrating covariates associated with each cell as inputs of its neural network. It can help condition the latent distributions and thus remove potential confounding factors. In particular, the batch information can be used as one conditioning covariate (among others). During the annotation learning process, Scyan corrects the batch effect by overlapping the inter-batch distributions on its latent space according to a chosen reference batch. We typically choose the batch reference for its representativeness and its large number of cells. As the network is invertible, we can map each cell from the latent space (in which there is no batch effect) to the original space as if all cells came from the reference batch. In brief, by using the same reference covariates for every cell, every variability due to different confounding factors (such as the batch) is removed.

\section*{Scyan latent space provides interpretability and helps population discovery}

Scyan's latent space is key for interpretability. Specifically, it enables the understanding of the Scyan annotation process, and also helps to quickly characterise new populations of cells to improve the annotation.
We illustrate population discovery on the POISED dataset. For this purpose, we show that we could annotate six populations that were missed during manual gating, such as differentiated effector T cells\cite{tem} (TCD8 TEM) and $\gamma \delta$TCR CD16+ cells. To demonstrate two different ways of discovering these populations, we show that we can (i) annotate more precise populations among known ones, or (ii) discover a population associated with cells that were left unlabelled (when there was no corresponding population from the biological table). One advantage of Scyan is its table flexibility: the new populations, once characterised, can be added to the knowledge table, and Scyan will then be able to annotate them.

\begin{figure*}[hbt!]
\centering
\includegraphics[width=\linewidth]{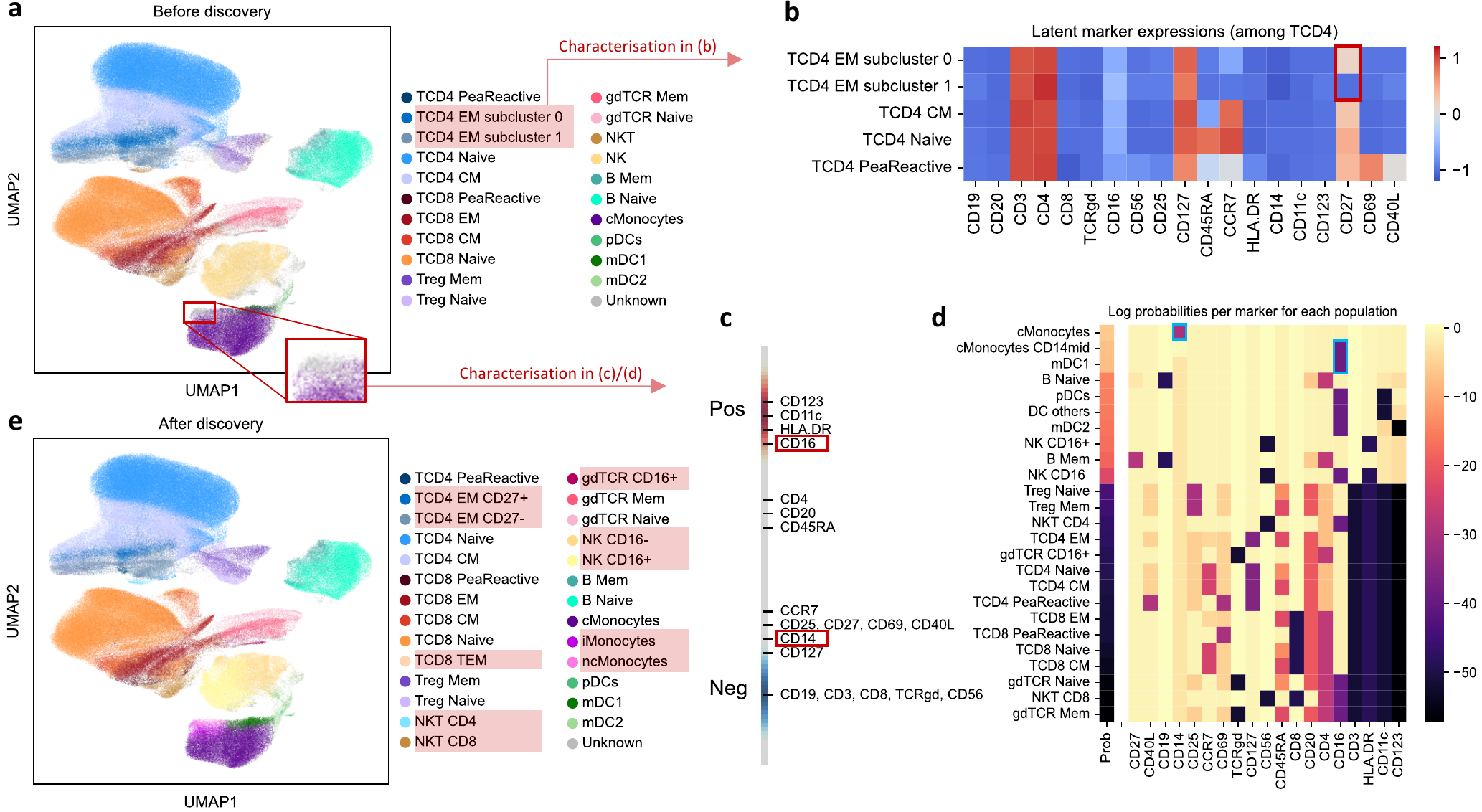}
\caption{\label{fig:4}\textbf{Interpretability and population discovery with Scyan.} \textbf{a}, UMAP on POISED before population discovery. Two subclusters of TCD4 EM cells have been defined and characterised in (b). Also, intermediate and non-classical monocytes were removed from the knowledge table to show that we can retrieve them: as shown by the red magnifying window, Scyan annotated these cells as unknown, and we characterise them in (c/d). \textbf{b}, Latent space expressions for subsets of TCD4 cells, displayed on a heatmap. We can easily see the difference between the two clusters: one is CD27+, the other CD27-. \textbf{c}, Scyan helps characterise the unknown cells defined in (a) by showing its latent marker expressions, displayed on a shared scale going from Negative to Positive expressions. \textbf{d}, Scyan provides soft predictions for all populations (first column), i.e. a log probability is associated with each population. Then, each population probability is decomposed into a sum of marker impact on the probability (one row). Dark colours indicate that the corresponding marker expression decreased the population probability of the corresponding row. For instance, the expression of CD14 (which is low for this population, according to (c)) decreased the confidence for predicting cMonocytes. \textbf{e}, UMAP of Scyan annotations after population discovery. The red boxes denote new populations compared to (a).}
\end{figure*}

\subparagraph{Understanding Scyan annotation process}

Scyan annotation process can be interpreted on one cell or a group of similar cells (see methods \autoref{interpretability}). Typically, we can select one population and interpret Scyan's annotation process on this group of cells. First, we can display all the latent marker expressions corresponding to these cells (\autoref{fig:4}c). It opens up a new simple way to understand which marker is positive or negative at a glance. Indeed, the latent space has a shared scale for all markers, and a simple scale indicates expression levels between Negative (-1) and Positive (+1). Then, the most important thing is the decomposition of Scyan confidence per population and per marker. Indeed, on \autoref{fig:4}d, we show that for each population, we can provide a confidence level, and each population confidence can be decomposed into a sum of confidence per marker. In brief, we show the importance and the impact of each marker on the prediction of each population. These confidences are quantified as log probabilities (i.e., negative values, and higher values indicate higher confidence). This interpretability property can be used to discover new populations (see \autoref{unknown_pops}).

\subparagraph{Annotating more precise populations}

When the defined populations are not precise enough, they can be split into new subpopulations. For this purpose, the first step is to subcluster populations: this is done using Leiden\cite{leiden} clustering. On \autoref{fig:4}a, we split T CD4 EM cells into two groups (subcluster 0 and subcluster 1). Then, we can characterise the subclusters using Scyan's latent space: \autoref{fig:4}b clearly shows that CD27 can explain the difference between the two (it is positive for one cluster and negative for the other). Traditionally, new subpopulations, once clustered, are characterised via either (i) back-gating (i.e., showing marker expressions of a subpopulation on multiple 2D scatter plots) or (ii) plotting a heatmap of expressions for all subpopulation. Yet, back-gating on 40 markers can be difficult and one may miss some important marker expressions. Also, while heatmaps of expressions retain global information, they may be difficult to read: using one common scale tends to erase the biological differences on markers whose expression scale is lower than for the others. Using Scyan latent space helps resolve this issue since all marker expressions can be described by their corresponding latent expressions, each having a similar range of values that improves readability. Afterwards (after characterisation), one can check that the marker expressions are correct using back-gating if needed.\\

As we have identified the two subclusters, we can then replace the "T CD4 EM" population by "T CD4 EM CD27+" and "T CD4 EM CD27-" in the knowledge table given to Scyan. Re-running the model leads to the annotation of these new populations (see \autoref{fig:4}e). This can be re-iterated multiple times, allowing us to go deeper into the sub-populations and, in the end, to understand the full diversity of cell types from the cytometry samples.\\

Note that one advantage over clustering is our computational efficiency. As clustering does not scale well (\autoref{fig:2}c), using a subset of cells is needed for characterisation, but Scyan, in the end, will annotate every cell and not just a subset.


\subparagraph{Annotating unknown populations}
\label{unknown_pops}
Sometimes, users may forget some populations in the table given to Scyan, and the corresponding cells will be left unclassified. Since every population from the POISED dataset was already described, we decided to remove two populations from the table provided to Scyan (non-classical and intermediate monocytes) to see if we could retrieve them. As shown in \autoref{fig:4}a on the red magnifying glass, cells corresponding to these populations were annotated as being "Unknown" (light grey color). We can further investigate these "Unknown" cells to retrieve their corresponding population. We can see, for instance, that these "Unknown" cells are CD16 positive and CD14 negative (\autoref{fig:4}c). Secondly, we show the confidence of Scyan in the prediction of all populations (\autoref{fig:4}d). According to the first column, we see that the first guesses of Scyan are classical monocytes (both CD14 high and mid) and mDC1. Then, we look at the three corresponding rows: they correspond to the confidence of Scyan for these populations decomposed by markers. For instance, we see that the expression of CD14 (which is negative, according to \autoref{fig:4}c) decreased Scyan confidence toward the prediction of classical monocytes. Thus, based on the first row, we can conclude that the 'Unknown' cells are similar to classical monocytes but are CD14- instead of CD14+, and that these cells are non-classical or intermediate monocytes. Similarly, the third row shows that they look like mDC1 cells but with a CD16+ expression instead of negative (again, \autoref{fig:4}c was needed to see the expression of CD16). Once more, we indeed conclude that these cells are non-classical/intermediate monocytes, and they can be added back to the table for the annotation (see \autoref{fig:4}e). To summarise, the process is the following: (i) we choose a group of cells that were unclassified by Scyan, (ii) we quickly characterise these cells using Scyan latent space, and (iii) we update the table to annotate them. Combining \autoref{fig:4}c and \autoref{fig:4}d provides a description of the Scyan annotation process that is understandable by humans, through decomposition into confidences by marker and by population.

\section*{Comparison to supervised models}

\subparagraph{Performances on POISED}

\begin{figure*}[hbt!]
\centering
\includegraphics[width=\linewidth]{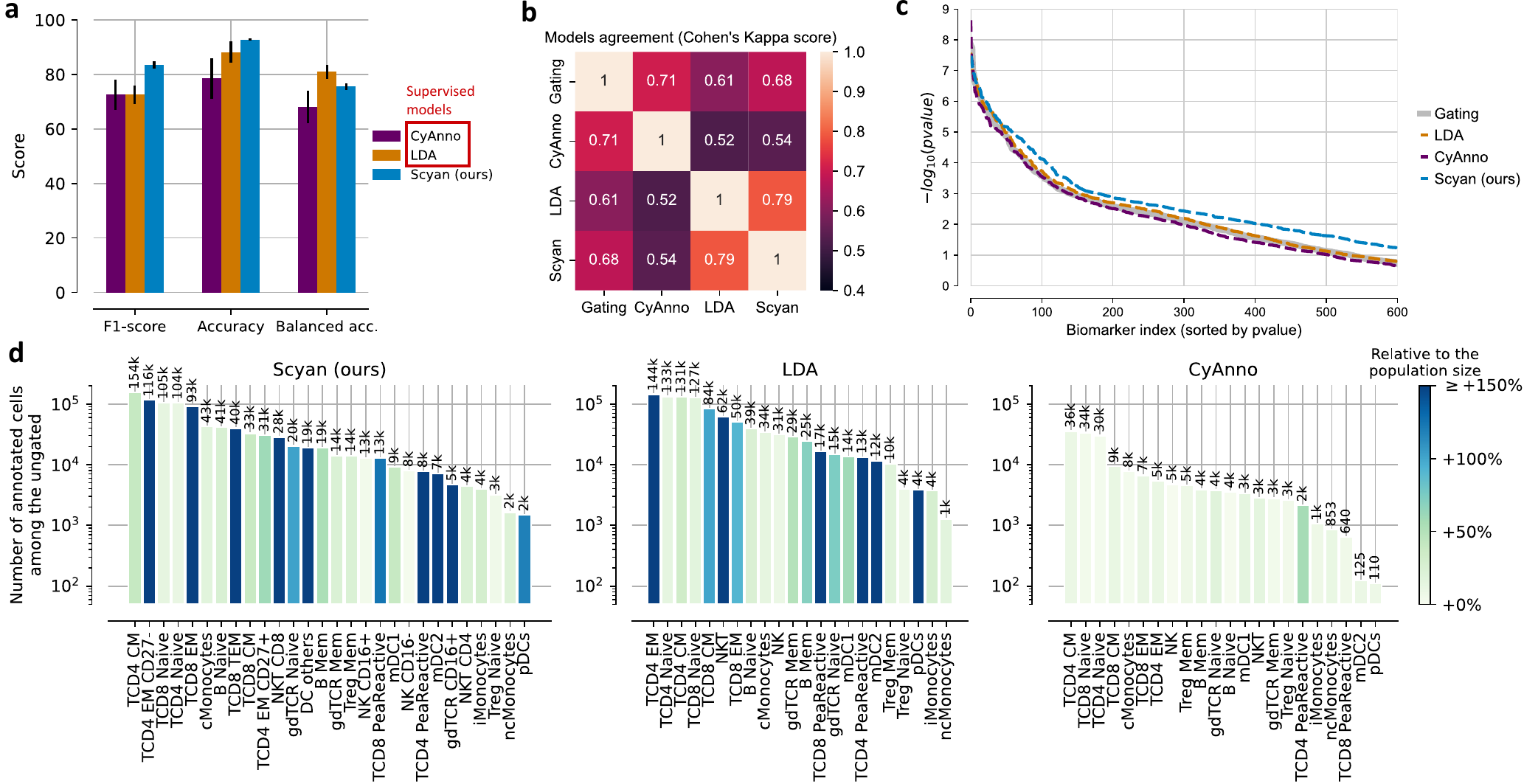}
\caption{\label{fig:5}\textbf{Comparison to supervised models.} The last two figures were done after Scyan's population discovery. \textbf{a}, Metrics on POISED. Note that among the three methods tested, CyAnno and LDA are supervised methods (i.e., using training labels from manual gating). \textbf{b} Heatmap representing pairwise models agreement using Cohen's Kappa score. A high value indicates a better agreement (the highest value is 1). \textbf{c}, After annotation, one can extract biomarkers and run differential expression relative to a clinical condition. Here, we show the significance of the biomarkers for all methods (higher is more significant). \textbf{d}, Number and percentage of cell types that were annotated by the models among the ungated ones.}
\end{figure*}

LDA\cite{lda} and CyAnno\cite{cyanno} (two supervised models) were compared to Scyan on the POISED dataset. For a realistic scenario, the two models were trained on one batch and evaluated on the others (one run for each batch). Scyan, in contrast, is unsupervised and can be run directly on all cells. \autoref{fig:5}a shows the performances of the three models on POISED. Even though the benchmark is in favor of supervised models (since they use labels), Scyan still has a higher performance. Also, since we are comparing an unsupervised method to supervised methods, comparing results to manual gating is biased. For this reason, we additionally compare the models' agreement (in a pairwise manner) using Cohen's Kappa score (\autoref{fig:5}b). We see that LDA and Scyan are the two models whose agreement is the highest, even higher than LDA and CyAnno (while they are two supervised methods trained for the same task). Concerning the disagreement between Scyan and the manual gating, we show in the supplementary \autoref{fig:sup_9}a,b that most disagreement is partly due to the subjective delimitation boundaries between non-classical/intermediate/classical monocytes. As shown by \autoref{fig:sup_9}ab, although Scyan is also properly annotating these populations, it has slightly different decision boundaries than manual gating, which still creates a significant decrease in F1-score or balanced accuracy. It emphasises again the importance of comparing the agreement between all models instead of only comparing to manual gating.

\subparagraph{Annotations of the ungated cells}
\label{ungated}

One key aspect of annotation models is whether they annotate more cells than traditional manual gating. This can enhance the biomarker discovery and provides higher statistical significance during post-annotation analyses. We compared the number of cells annotated by Scyan, LDA \cite{lda}, and CyAnno \cite{cyanno} on POISED, and demonstrated that Scyan annotates more cells than CyAnno, and a similar amount of cells to LDA (\autoref{fig:5}d). Indeed, CyAnno annotated 15\% of the ungated cells, Scyan 97\%, and LDA was set up to annotate all cells. Moreover, Scyan annotated 6 more populations compared to CyAnno and LDA. Most importantly, we show by back gating that the annotated cells were properly classified (see \autoref{fig:sup_9}c). Indeed, one limitation of supervised models such as LDA or CyAnno is that they can not annotate new populations, i.e., they are limited to populations that were manually gated. Although knowledge-based annotation models (like ours) are limited to populations from the provided table, the table can be easily extended. This property is, therefore, crucial for population discovery with Scyan.

\subparagraph{Usage for biomarker discovery}

The POISED dataset is decomposed into two conditions: peanut-stimulated samples, and unstimulated ones. We try to find biomarkers that are differentially expressed on peanut-stimulated samples. For that, for all models, biomarkers were extracted, and we ran paired-T-tests between the two conditions. On \autoref{fig:5}c, we sorted the biomarkers by p-value for all models, and we display the $-log_{10}(p-value)$ of the first 400 biomarkers. We show that Scyan extracts more biomarkers of higher significance. Note that a similar process could be run for other clinical conditions such as the patient response to treatment. Having more significant biomarkers means that it will be easier to better predict such an outcome.

%% file: 3_discussion.tex
\section{Discussion}
We have introduced Scyan, a multitask neural network for cytometry annotation, batch-effect removal, debarcoding, and population discovery. It provides a robust and broad pipeline to analyse cytometry cell populations, monitor their dynamics over time, and compare the populations' proportions among patients. Such analyses can help discover biomarkers or specific populations characteristic of response to treatment, for example. 
Scyan can perform fast and automatic annotations for these large datasets and correct potential batch effects. Some studies use barcoding to reduce the batch effect, hence requiring a debarcoding step that Scyan can also perform. Thus, Scyan is suitable for various types of cytometry projects and does not rely on any extra cytometry analysis library.

Scyan annotates populations without needing labels and, therefore, can fully replace manual gating. It uses a marker-population table containing expert knowledge. The literature offers many resources and existing knowledge to construct such tables, but some marker expressions remain unexplored. For this reason, we offer the possibility to handle "not applicable" values inside the table and, to improve flexibility, intermediate expressions such as "mid" or "low". In the case where the panel remains not well known enough to build the input table, Scyan can help discover new populations: analyses start by annotating large populations and then gradually target smaller and smaller cell types. Also, with the increasing usage of cytometry, we expect the marker knowledge to improve over time, reinforcing Scyan performance and ease of use.

In terms of model architecture, normalizing flows are a recent and promising field of research in generative models. They benefit from interesting mathematical properties such as (i) exact likelihood computation and (ii) invertibility. We show that normalizing flows can be used to leverage marker knowledge in a biologically natural way, providing interpretability. Indeed, the network invertibility allows switching between the measured marker expressions and their latent expressions. In this space, all latent markers have unified expression ranges, which is convenient for human analysis, especially for population discovery. It also makes the model reliable and transparent to biologists, which can help build trust toward the model annotations and validate them. 
Moreover, normalizing flows are smooth transformations that control how the space is deformed, ensuring that we do not alter the biological meaning behind marker expressions. At the same time, it benefits from the expressiveness and flexibility of deep neural networks. In fact, the usage of neural networks allows adding additional terms in the loss function to handle the batch effect, which is naturally corrected with the network invertibility. Eventually, we can further push the usage of these convenient mathematical properties for other tasks in single-cell analysis, for instance, single-cell RNA sequencing data or imaging mass cytometry data\cite{imc}.

%% file: 4_methods.tex
\section{Online Methods}

\section*{Scyan model}

In this section, we formulate the annotation problem and detail the Scyan model illustrated in \autoref{fig:1}b/c. Let $\vec{x_1},  \dots, \vec{x_N} \in \mathbb{R}^M$ represent the vectors of $M$ marker expressions for $N$ cells. We assume these expression levels have already been transformed using the $asinh$ or logicle\cite{logicle} transformation and standardised. Our objective is to associate each cell to one of the $P$ predefined cell types using a marker-population table $\vec{\rho} \in \mathbb{R}^{P \times M}$, with $\rho_{z,m}$ summarising the knowledge about the expression of marker $m$ for population $z$. If it is known that population $z$ expresses $m$ then $\rho_{z,m} = 1$; if we know that it does not express $m$ then  $\rho_{z,m} = -1$. Otherwise, if we have no knowledge or if the expression can vary among the population, then  $\rho_{z,m} =\mbox{NA}$. Note that it is also possible to choose values in $\mathbb{R}$; for instance, for mid or low expressions, we can choose $0$ and $0.5$ respectively (see Supplemental \autoref{supp_advice}). In addition, we can add covariates $\vec{c_1}, \dots, \vec{c_N} \in \mathbb{R}^{M_c}$ associated with each cell, e.g., information about the batch or which antibody has been used by the cytometer. $M_c$ denotes the number of covariates; it can be zero if no covariate is provided.  

\subparagraph{Generative process}

Let $\vec{X}$ be the random vector of size $M$ representing one cell by its standardised marker expressions; in other words, $\vec{X}$ is the random variable from which $\vec{x_1}, \dots, \vec{x_N}$ are sampled. We model $\vec{X}$ by the following deep generative process:

\begin{equation}
  \begin{gathered}
    Z \sim Categorical(\vec{\pi}) \\
    \vec{E} \; | \; Z = (e_m)_{1 \leq m \leq M} \mbox{, where }
        \left\{
            \begin{array}{ll}
                e_m = \rho_{Z,m} & \mbox{if }\rho_{Z,m} \neq \mbox{NA} \\
                e_m \sim \mathcal{U}([-1, 1]) & \mbox{otherwise,}
            \end{array}
        \right.\\
    \vec{H} \sim \mathcal{N}(\vec{0}, \sigma \mathbb{\vec{I_M}}) \\
    \vec{U} = \vec{E} + \vec{H} \\
    \vec{X} = f_{\vec{\phi}}^{-1}(\vec{U}).
  \end{gathered}
\end{equation}

In the above equations, $\vec{\pi} = (\pi_z)_{1 \leq z \leq P}$ represents the weights of each population, with the constraints $\pi_z \geq 0$ and $\sum_z \pi_z = 1$. $Z$ is the random variable corresponding to a cell type among the $P$ possible ones. $\vec{E}$ is a population-specific variable whose terms are either known according to the expert knowledge table $\vec{\rho}$ or drawn from a uniform distribution between negative expressions (represented by -1) and positive expressions (represented by +1). $\vec{H}$ contains cell-specific terms, such as autofluorescence. Finally, $\vec{U}$ is the cell's latent expressions, summing a population-specific component and a cell-specific one. Also, $\vec{U}$ can be transformed into a measured cell marker expressions vector $\vec{X}$ by the inverse of a deep invertible network $f_{\vec{\phi}}$ detailed below.

\subparagraph{Invertible transformation network}

The core network, $f_{\vec{\phi}}$ (illustrated in \autoref{fig:1}b), is a normalizing flow\cite{nf1, nf2, nf3}. It transforms the target distribution $p_X$ into the known base distribution $p_U$, which was described in the previous section. Using a change of variables, we can compute the exact likelihood of a sample $\vec{x}$ by:
\begin{equation}
\label{eq:cv}
p_X(\vec{x}; \vec{\theta}) = p_U(f_{\vec{\phi}}(\vec{x}); \vec{\pi}) \cdot log \;  \Big| det \frac{\partial f_{\Vec{\phi}}(\Vec{x})}{\partial \Vec{x}^T} \Big|.
\end{equation}
To be able to compute this expression, we need to choose an invertible network with a tractable Jacobian determinant. We have chosen a set of transformations called Real Non-Volume-Preserving (Real NVP\cite{realnvp}) transformations, which are compositions of functions, named coupling layers $f_{\Vec{\phi}} := f^{(L)} \circ f^{(L-1)} \circ \dots \circ f^{(1)}$ with $L$ the number of coupling layers. Each coupling layer $f^{(i)}: (\Vec{x}, \Vec{c}) \mapsto \Vec{y}$ splits both $\Vec{x}$ and $\Vec{y}$ into two components $(\Vec{x^{(1)}}, \Vec{x^{(2)}}), (\Vec{y^{(1)}}, \Vec{y^{(2)}})$ on which distinct transformations are applied. We propose below an extension of the traditional coupling layer\cite{realnvp} to integrate covariates $\vec{c}$ (illustrated in \autoref{fig:1}c): 

\begin{equation}
    \begin{cases}
      \Vec{y^{(1)}} = \Vec{x^{(1)}}\\
      \Vec{y^{(2)}} = \Vec{x^{(2)}} \odot exp\Big(s([\Vec{x^{(1)}}; \Vec{c}])\Big) + t([\Vec{x^{(1)}}; \Vec{c}]).
    \end{cases}       
\end{equation}

In the equations above, $\odot$ stands for the element-wise product, $[.;.]$ is the concatenation operator, and $(s, t)$ are functions from $\mathbb{R}^{d + M_c}$ to $\mathbb{R}^{M-d}$ where $d$ is the size of $\Vec{x^{(1)}}$. These functions can be arbitrarily complex, in our case, multi-layer-perceptrons. Note that the indices used by the coupling layer to split $\Vec{x}$ into $(\Vec{x^{(1)}}, \Vec{x^{(2)}})$ are set before training and are different for every coupling layer. This way, we ensure that the flow transforms all the markers. Each coupling layer has an easy-to-compute log Jacobian determinant, which is $\sum_i s([\Vec{x^{(1)}}; \Vec{c}])_i$, and is easily invertible as shown in the following equations:

\begin{equation}
    \begin{cases}
      \Vec{x^{(1)}} = \Vec{y^{(1)}}\\
      \Vec{x^{(2)}} = (\Vec{y^{(2)}} - t([\Vec{y^{(1)}}; \Vec{c}])) \odot exp\Big(-s([\Vec{y^{(1)}}; \Vec{c}])\Big).
    \end{cases}       
\end{equation}

As $f_{\Vec{\phi}}$ is a stack of coupling layers, it is also invertible, and its log Jacobian determinant is obtained by summing each coupling layer log Jacobian determinant. Stacking many coupling layers is essential to learning a rich target distribution and complex variables interdependencies. Overall, the normalizing flow has some interesting properties: (i) the coupling layers preserve order relation for two different expression values, and (ii) penalise huge space distortion (the log determinant term). The two properties are useful to preserve the biological variability as much as possible.

\subparagraph{Learning process}

The model parameters are $\vec{\theta} = (\vec{\pi}, \vec{\phi})$. For computational stability during training, instead of learning $\vec{\pi}$ itself we actually learn logits $(l_z)_{1 \leq z \leq P}$ from which we obtain $\pi_z = \frac{e^{l_z}}{\sum_k e^{l_k}}$. By doing this, we ensure the positivity of each weight and guarantee they sum to 1. To train the model, we minimise the Kullback–Leibler (KL) divergence between the cell's empirical marker-expression distribution $p_{X^*}$ and our model distribution $p_X$. It is equivalent to minimising the negative log-likelihood of the observed cell expressions $-\mathbb{E}_{\Vec{x} \sim p_{X^*}}\Big[ log \; p_X(\Vec{x}; \vec{\theta}) \Big]$ over $\Vec{\theta}$. Using \autoref{eq:cv} and adapting it to integrate covariates leads to minimising the following quantity:

\begin{equation}
    \mathcal{L}_{KL}(\vec{\theta}) = - \sum_{1 \leq i \leq N} \bigg[ log \: \Big( p_U(f_{\vec{\phi}}(\vec{x_i}, \vec{c_i}); \vec{\pi}) \Big) + log \;  \Big| det \frac{\partial f_{\Vec{\phi}}(\Vec{x_i}, \vec{c_i})}{\partial \Vec{x}^T} \Big| \bigg].
\end{equation}

 In the above equation, $p_U(f_{\vec{\phi}}(\vec{x_i}, \vec{c_i}); \vec{\pi}) = \sum_{z=1}^P \pi_z \cdot p_{U \mid Z = z}(f_{\vec{\phi}}(\vec{x_i}, \vec{c_i}))$, which is not computationally tractable because the presence of NA in $\vec{\rho}$ leads to the summation of a uniform and a normal random variable. We approximate the density of the sum of the two random variables by a piecewise density function that is constant on $[-1 + \sigma, 1 - \sigma]$ with Gaussian queues outside of this interval. In practice, we choose a normal law with a low standard deviation, which leads to a good piecewise approximation (see Supplemental \autoref{supp_approx}). If we consider the KL-divergence as described above, some modes may collapse; that is, one small population may not be predicted. Indeed, a small population $z$ that has a small weight $\pi_z$ leads to smaller gradients towards this population. To solve this issue, we favour small populations once every two epochs. For that, for all $z$, we replace $\pi_z$ by $\pi_z^{(-T)} = \frac{e^{-l_z / T}}{\sum_k e^{-l_k / T}}$ where $T$ is called temperature\cite{temp1, temp2} as it increases the entropy of $\vec{\pi}^{(-T)}$. Note that here we added the minus signs to reverse the weights of the populations so that it favours small ones. A temperature close to 0 leads to high weights for small populations, while an infinite temperature leads to equal population weights, i.e., the maximum entropy. Alternating between $\vec{\pi}$ and $\vec{\pi}^{(-T)}$ allows for a better balance of population sizes at the end of the training.

We optimize the loss on mini-batches of cells using the Adam optimizer\cite{adam}. Once finished training, the annotation process $\mathcal{A}_{\vec{\theta}}$ consists in choosing the most likely population according to the data using Bayes's rule. So, for a cell $\vec{x}$ with covariates $\vec{c}$, we have:
\begin{equation}
\label{eq:a}
\mathcal{A}_{\vec{\theta}}(\vec{x}, \vec{c}) = \argmax_{1 \leq z \leq P} \; \; \pi_z \cdot p_{U \mid Z = z}(f_{\vec{\phi}}(\vec{x}, \vec{c})).
\end{equation}

We also define a log threshold $t_{min}$ to decide whether or not to label a cell, i.e., we don't label a cell if: $$max_{1 \leq z \leq P} \; \; p_{U \mid Z = z}(f_{\vec{\phi}}(\vec{x}, \vec{c})) \leq e^{t_{min}}$$

\section*{Batch-effect correction}

When the batches are provided into the covariates, the normalizing flow will naturally learn to align the latent representations of the multiple different batches. Using this property, combined with the network invertibility, enables batch-effect correction. To effectively correct the batch effect of a sample $\vec{x}$ with covariates $\vec{c} \neq \vec{c_{ref}}$, we first transform $\vec{x}$ into its latent expressions via $f_{\vec{\phi}}$. Since the latent space is batch-effect free, latent expressions can then be transformed back into the original space using the covariates of the reference batch and $f_{\vec{\phi}}^{-1}$. Formally, we denote by $\tilde{\vec{x}}$ the batch-effect corrected cell associated to $\vec{x}$; that is, $\tilde{\vec{x}} = f_{\vec{\phi}}^{-1}\Big(f_{\vec{\phi}}(\vec{x}, \vec{c}), \vec{c_{ref}}\Big)$. In this manner, we get expressions $\tilde{\vec{x}}$ as if $\vec{x}$ were cell expressions from the reference batch. Thus, the latent space is used to align the distributions in the original space. Note that, since batch-effect was corrected in the original space, it is possible to reuse a trained dimension reduction tool (e.g., a UMAP). Applying this UMAP again does not change the dimension reduction of the cells from the reference (due to the network invertibility), but it updates those of all the other batches. Note that batch-effect is corrected on the markers provided to Scyan. Thus, for markers not provided to Scyan, the batch correction will not be applied. For this reason, one can provide all markers from the panel, even if it means having some columns with "NA" only.

\section*{Interpretability and population discovery}

\subparagraph{Understanding Scyan predictions}
\label{interpretability}
One important thing to notice is that $U_1 \; | \; (Z = z), \dots, U_M \; | \; (Z = z)$ are independent for every population $z$. It means that we can decompose $log \; p_{U \mid Z = z}(\vec{u}) = \sum_m log \; p_{U_m \mid Z = z}(u_m)$, and we can gather all these terms into a matrix of scores $\Big(log \; p_{U_m \mid Z = z}(u_m)\Big)_{z,m}$. The term $log \; p_{U_m \mid Z = z}(u_m)$ can be interpreted as the impact of marker $m$ towards the prediction of the population $z$ for the latent cell expression $\vec{u}$. Based on that, we can interpret Scyan predictions for a group of cells $(\vec{x_i}, \vec{c_i})_i$ by transforming the cells into their latent expressions and then averaging the score matrices. The resulting matrix is typically displayed on a heatmap (\autoref{fig:4}d), and populations are sorted by their score (sum over a score matrix row). Note that, in the figure, each population score is scaled to make it easier to read.

\subparagraph{Latent expressions}
Scyan interpretability is based on its latent space. Considering a cell $\vec{x}$ and its covariates $\vec{c}$, its latent representation is $\vec{u} = f_{\vec{\phi}}(\vec{x}, \vec{c})$. The information of which marker is positive or negative is contained in $\vec{u}$. Indeed, $u_m \approx 1$ corresponds to a positive expression, while $u_m \approx -1$ represents a negative expression, whatever the marker $m$ (i.e., expression levels for all markers are unified). Similarly, $u_m \approx 0$ is a mid-expression, and so on. We average the latent cell expressions over one population to obtain a latent expression at the population level. These population-level latent expressions can be displayed for one population (\autoref{fig:4}c) or for all of them at once (\autoref{fig:4}b).

\subparagraph{Differential expressions for population discovery}
We extend this interpretability tool for population discovery. We first perform a Leiden clustering\cite{leiden} over all the cells with a high resolution to get a large number of clusters. Then, we use this clustering to separate each predicted population into multiple sub-clusters. Each sub-cluster can then be explored using the previously described latent space, showing exactly where the differences between each sub-cluster are. On \autoref{fig:4}b/c, the model automatically fills unknown expressions (NA in the table) since the latent space represents expressions for all markers. Also, if some protein markers of the panels are entirely unknown, they can still be added inside the marker-population table as a column of "NA". Even though such a marker does not provide much information for the prediction, it can be very informative for population discovery as it will thus appear in the heatmap. Also, note that some clusters appear as "NA" in \autoref{fig:4}f. They correspond to sub-clusters that were not significant enough according to a threshold in terms of the ratio of cells. If too few clusters are displayed, one can increase the resolution to obtain more clusters.

\subparagraph{Automatic scatter plots}
We automatically choose a set of very discriminative markers to enhance two-dimensional scatter plots. For that, let $K$ be the number of markers requested by the user. For each marker of the panel, we compute the sum of the Kolmogorov-Smirnov test for every population of interest in a "one-versus-all" manner (the user also chooses the populations of interest). Each marker has its own statistic, and we choose the $K$ first markers whose statistics are the highest.

\section*{Model hyperparameter optimisation}

One important issue in training deep learning models is fine-tuning their hyperparameters. Because our model is unsupervised, we cannot consider any supervised metric such as accuracy. We thus have to use an unsupervised metric that measures the annotation quality. For this reason, we defined a heuristic that combines (i) a clustering metric, the Davies-Bouldin Index\cite{dbs} (DBI), to obtain well-separated clusters, (ii) a count of the missing populations to favour the presence of all populations among the predictions, (iii) a Dirichlet probability on population weights to favour population diversity, and (iv) the iLISI score. Formally, let $O$ be the number of populations that the model did not predict at all, and $\vec{X}, \vec{y}_{pred}$ the cells' expressions and predictions, respectively. Then, we define our heuristic to be minimised by $(O + 1) \cdot DBI(\vec{X}, \vec{y}_{pred}) \cdot (- \sum_z log \; \pi_z) / iLISI(\vec{X}, \vec{y}_{pred})$. Note that $- \sum_z log \; \pi_z$ is proportional to the log Dirichlet probability of the learned population weights $\vec{\pi}$. An advantage of using the DBI is that it is computationally more efficient than some clustering metrics, such as the silhouette score\cite{silhouette}. In particular, the DBI scales efficiently to large datasets. Also, if batch covariates were not provided, we simply remove the iLISI term in the heuristic definition.

\section*{Benchmark-related methods}

\subparagraph{Metrics definition}

Most metrics of the benchmark are implemented in Scikit-learn\cite{sklearn}: the accuracy, the balanced accuracy, the macro-averaged F1-score, the Davies Bouldin Score, Cohen's Kappa score, the Silhouette score, and the Calinski Harabasz Score. Only the cLISI\cite{harmony} metric was not implemented in Scikit-learn but we used the one from Harmony\cite{harmony}. For one cell, the cLISI is the number of different cell types that can be found in the close neighbourhood of the cell. Thus, a lower value indicates a better cell-type separation, and the best value is 1. We also used the iLISI\cite{harmony} metric for the hyperoptimisation of our model. For one cell, it corresponds to the number of different batches that can be found in the close neighbourhood of the cell. Thus, a higher value indicates a better batch correction, and the best value is the number of different batches. For the batch correction benchmark, evaluation was run on the cell-type related markers only: CD19, CD20, CD3, CD4, CD8, TCRgd, CD16, CD56, CD25, CD127, CD45RA, CCR7, HLA.DR, CD14, CD11c, CD123, CD27, CD69, CD40L.

\subparagraph{Batch effect amplification}
\label{bea}
On the POISED dataset, we amplified the batch effect so that the benchmark becomes more complex. Let $\sigma \geq 0$ a scale factor, and $b_1, \dots, b_N \in [1 \dots 7] $ the batch number associated to each of the $N$ cells. Then, we sample 7 matrices $\vec{S}_1, \dots, \vec{S}_7 \in \mathbb{R}^{M \times M}$, whose elements are drawn from $\mathcal{N}(0, \sigma)$. For a cell $i$ of expression $\vec{x}_i$, the batch-effect-amplified expression $\vec{x}_i^{'}$ is multiplied by some batch-relative term: $\vec{x}_i^{'} = (\vec{I}_M + \vec{S}_{b_i})\vec{x}_i$. In this equation, $\vec{I}_M$ is the identity matrix of size $M \times M$, and the multiplication operation is matrix multiplication. In practice, we use $\sigma = 0.01$. Note that, the UMAPs were computed on the cell-type related markers, and did not use cell-state markers.

\subparagraph{Data preprocessing}
To compare our model with the other methods, we used a similar data preprocessing, and the same knowledge tables as MP\cite{mp} and ACDC\cite{acdc} for the AML, BMMC, and debarcoding datasets. On POISED, due to the impossibility of having non-binary values in the input tables of MP and ACDC, intermediate Monocytes had to be removed. Also, ACDC and MP used $x \mapsto asinh(\frac{x-1}{5})$ to preprocess marker expressions, while we used $x \mapsto asinh(\frac{x}{5})$. Note that we also standardised the data (required to run Scyan). Concerning the debarcoding task, we used the logicle transformation\cite{logicle} to preprocess marker expressions and then standardisation.

\subparagraph{Implementation details}
We implemented our model using Python and the Deep Learning framework Pytorch\cite{pytorch}. We used between 6 and 8 coupling layers whose multi-layer-perceptrons ($s, t$) have each between 6 and 8 hidden layers depending on hyperparameter optimisation. The hidden layer size can vary between 16 and 32.

%% file: 6_supp.tex
\section*{Supplementary information}

\subparagraph{Approach justification and related work}
In cytometry, when it comes to model the probability density functions of multidimensional marker expressions, their appearances make it natural to first consider Gaussian Mixture Models (GMM). However, in practice, each component of a GMM estimated from the data may not necessarily map to one population. Indeed, two small populations can be merged into one, and one large population may be split into two components with no interesting biological distinction between them. Also, we would have to annotate each component of the mixture. It could be done manually or using a semi-supervised approach. Yet, as discussed in the introduction, we prefer to use only the knowledge table $\Vec{\rho}$ instead. In terms of deep generative models, there are two main reasons to choose a Real NVP (the normalizing flow architecture) over GANs\cite{gan}, and VAEs\cite{vae}: the flow invertibility and the ability to compute the exact likelihood of a sample. Indeed, the flow invertibility enables a natural and simple way to correct the batch effect by transforming latent expressions back into the original space. Moreover, the ability to compute the exact likelihood of samples makes the annotation straightforward using the Bayes rule and the known base distribution. Another interesting property is that the Real NVP has a triangular Jacobian with positive terms on the diagonal. It enforces the model to diffuse the marker expressions slowly and prevents multiplication by a negative term. Indeed, such smooth transformations are essential to ensure that we do not mix the mapping between a population component and its actual marker expressions density. Also, the Jacobian determinant term in the loss function controls how much the flow dilates volumes in a point neighbourhood. This term thus prevents the collapse of a vast part of the space into a tiny component of the base distribution. From a biological point of view, the Real NVP transformations can be seen as compositions of complex compensations and monotonic transformations learned via deep learning.

\begin{table}
    \caption{Models properties comparison. We listed all the models considered in this article as well as manual gating.}
    \label{tab:supp_table}
    \centering
    \begin{tabular}{l c c c c c c c c}
        \toprule
    &   {\thead{\rotatebox{90}{Manual Gating}}} &   {\thead{\rotatebox{90}{Baseline}}}  &  {\thead{\rotatebox{90}{Phenograph}}}  & {\thead{\rotatebox{90}{MP}}} &   {\thead{\rotatebox{90}{ACDC}}}   &   {\thead{\rotatebox{90}{LDA}}} & {\thead{\rotatebox{90}{CyAnno}}} & {\thead{\rotatebox{90}{Scyan (ours)}}}  \\
    \midrule
Reproductible &  & X & X & X & X & X & X & X \\
Does not rely on manual gating & & X & X & X & X & & & X \\
Population discovery & X &  & X &  & X & & & X \\
Soft predictions &  & X &  &  & X & X & X & X\\
Interpretable & X &  & X & X & & &  & X \\
Generative model &  &  &  &  &  & & & X \\
Batch-effect correction &  &  &  &  &  & & & X \\
        \bottomrule
    \end{tabular}
\end{table}

\subparagraph{Density approximation in the presence of NA}
\label{supp_approx}

\begin{figure*}[hbt!]
\centering
\includegraphics[width=0.7\linewidth]{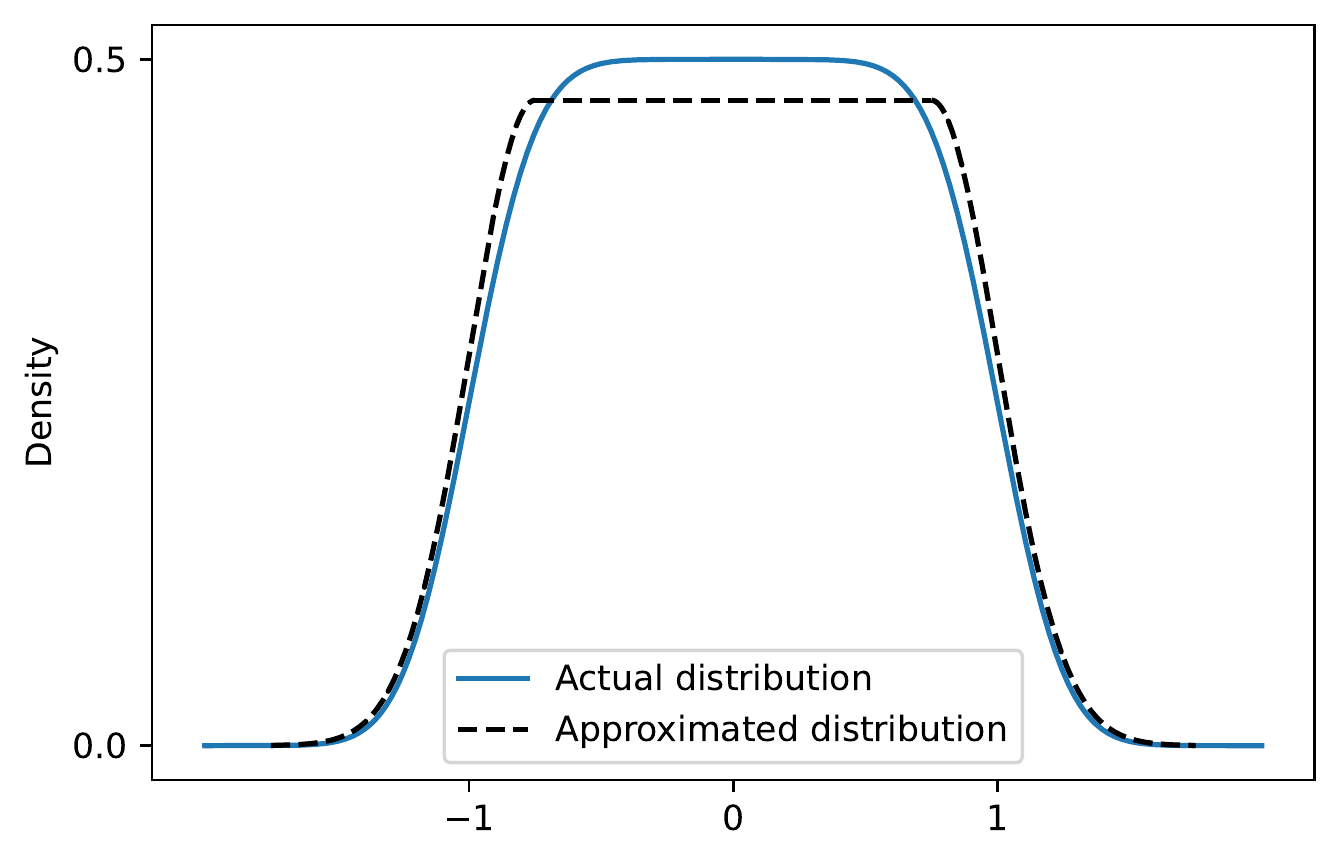}
\caption{\label{fig:approx}\textbf{Density approximation in the presence of NA.} If $z$ and $m$ are a population and a marker respectively such that $\rho_{z,m} = \mbox{NA}$, then $p_{U_m \; | Z = z}$ is approximated. This figure illustrates the actual and the approximated distribution with $\sigma = \frac{1}{4}$.}
\end{figure*}

Let $z$ be a population and $m$ a marker such that $\rho_{z,m} = \mbox{NA}$. It leads to $E_m \; | \; (Z = z) \sim \mathcal{U}([-1, 1])$ and $H_m \sim \mathcal{N}(0, \sigma)$. Thus, $U_m = E_m + H_m$ does not have a simple density expression. We approximate the probability density function $p_{U_m \; | Z = z}$ by the following function (see \autoref{fig:approx}), where $r = 1 - \sigma$ and $\gamma = \frac{\sigma \sqrt{2\pi}}{2r + \sigma \sqrt{2\pi}}$:
\begin{equation}
    \tilde{p}_{U_m \; | Z = z}(u) =
        \left\{
            \begin{array}{ll}
                \gamma \cdot \mathcal{N}(u + r;0, \sigma) & \mbox{if}\; u \leq -r\\
                \frac{\gamma}{\sigma\sqrt{2\pi}} & \mbox{if}\; u \in ]-r, r[\\
                \gamma \cdot \mathcal{N}(u - r;0, \sigma) & \mbox{if}\; u \geq r.
            \end{array}
        \right.\\
\end{equation}

Gradients are null in $]-r, r[$, and the queues of the approximated distribution are similar to the actual one. This expression is easy to compute, efficient during training, and a close approximation.

\subparagraph{Details on the knowledge table to be provided}
\label{table_details}

Scyan requires a table, a.k.a. biological prior table or knowledge table, to describe the populations to be annotated. The table is of size $P$ x $M$, where $P$ is the number of populations and $M$ the number of markers. For a given population $z$ and a marker $m$, the value at the row $z$ and the column $m$ of the table has to be one of the following:

\begin{itemize}
    \item $1$ if the population $z$ is known to express marker $m$
    \item $-1$ if the population $z$ is known not to express marker $m$
    \item $NA$ if we don't know anything about the expression of $m$ on population $z$
    \item a value in $]-1, 1[$ if the expression is known, but not positive or negative. For instance, $-0.5$ can be chosen for a low expression, and $0$ for a mid expression. It allows to better annotate complex populations, in particular population continuums (e.g., classical/intermediate/non-classical monocytes).
\end{itemize}

\subparagraph{Advice to build the marker-population table}
\label{supp_advice}

The design of the knowledge table is essential for the annotations. The literature can help its creation, but we also provide some advice to enhance the table:
\begin{itemize}
    \item It is better to provide no knowledge than false information; therefore, the user should feel comfortable using "Not Applicable" for a marker when unsure. Besides, if needed, population discovery can be used to go deeper into this marker afterwards.
    \item Note that the model interprets NA values by "any expression is possible". Thus, a population described with extensive use of NA values (e.g., above 90\% of markers, with no discriminative marker provided) can be over-predicted. This is a normal behaviour since few constraints are applied to this population.
    \item We enable the usage of intermediate expressions such as "mid" and "low" in the table. Yet, we advise using it only to differentiate two similar populations. Overusing these intermediate expressions in the table will challenge the user to create the table properly while not improving the results.
    \item It is not required to use all the panel markers. If some markers are unimportant for the annotation, they can be removed from the knowledge table.
\end{itemize}

\begin{figure*}[hbt!]
\centering
\includegraphics[width=\linewidth]{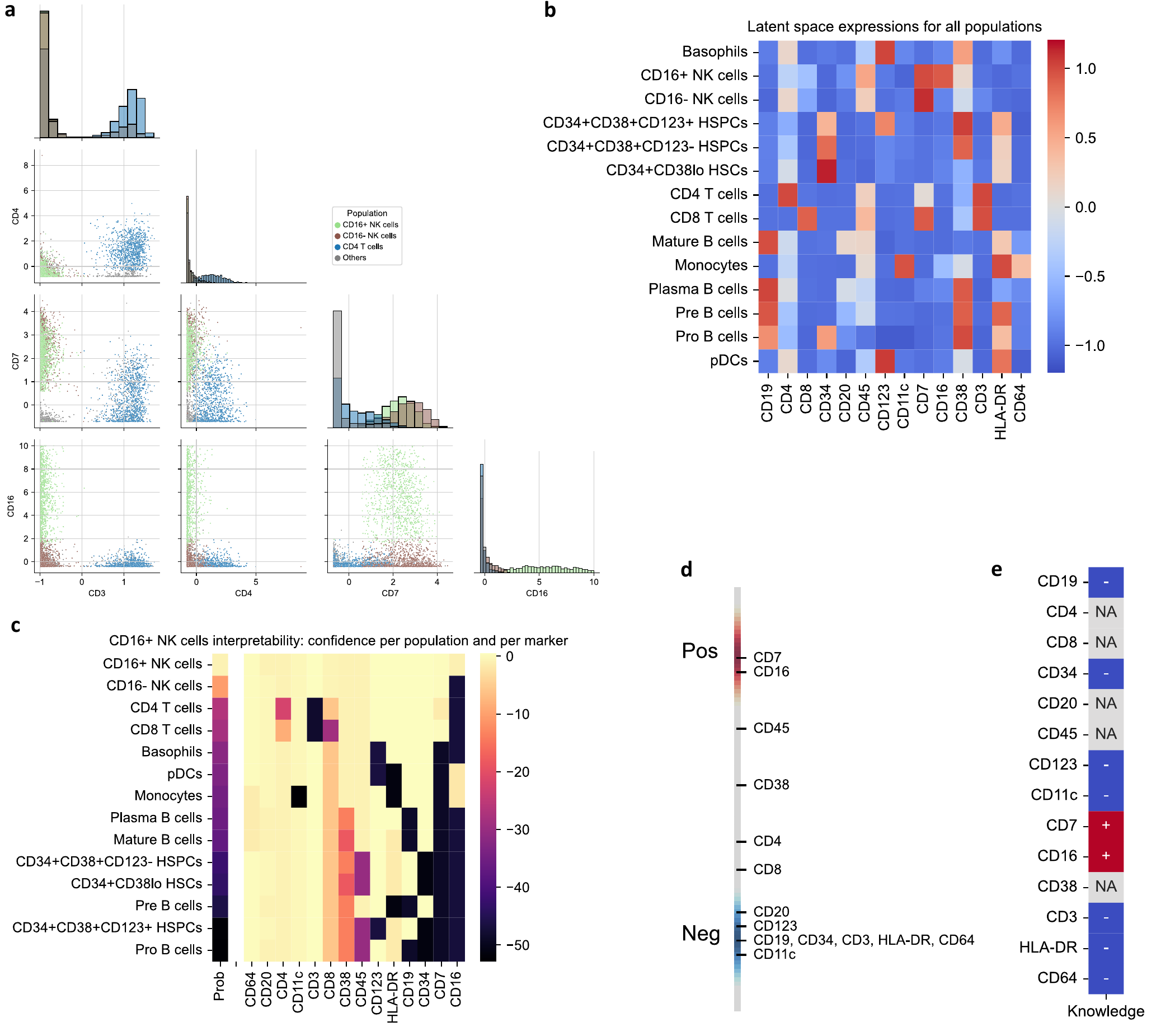}
\caption{\label{fig:sup_5}\textbf{Scyan visualisation, interpretability, and discovery on AML.} \textbf{a}, Separation of the CD4 T cells and the two NK populations on multiple scatter plots using CD3, CD4, CD7 and CD16 (standardised marker expressions). \textbf{b}, Scyan latent space for all populations. The latent space comprises one value per marker whose typical range is [-1, 1]. The closer to -1, the more negative the marker expression, and the closer to 1, the more positive the marker expression. \textbf{c}, Understanding Scyan predictions for CD16+ NK cells by providing Scyan confidence (or probability) for each population, each of them decomposed per marker. \textbf{d}, Scyan latent space for CD16+ NK cells, in other words, their expressions for all the considered markers. \textbf{e}, Extract of the knowledge table concerning CD16+ NK cells. Some markers were known to be positive, others negative, and some marker expressions were unknown or not applicable (NA).}
\end{figure*}

\subparagraph{Datasets details}
\label{supp_dataset}

For the AML dataset, we use 14 markers of the panel to identify 14 populations, while for the BMMC dataset, we use 13 markers to identify 19 populations. Concerning the POISED dataset, 19 markers were used to annotate 24 populations.

\subparagraph{Comments on supervised models runtime}

Since LDA and CyAnno rely on manual gating, the total time needed for the annotation is highly dependent on the time required for the manual annotation. For complex datasets with many patients and batch effect, manual gating can take full days to complete a precise annotation and thus to be able to run the supervised models. Also, if a population was discovered afterwards, manual gating has to be modified to target this new population. On the opposite, adding a new population to Scyan’s table (basically, modifying one or two marker expressions) is all we need to re-run the algorithm and target this new population.

\begin{figure*}[hbt!]
\centering
\includegraphics[width=\linewidth]{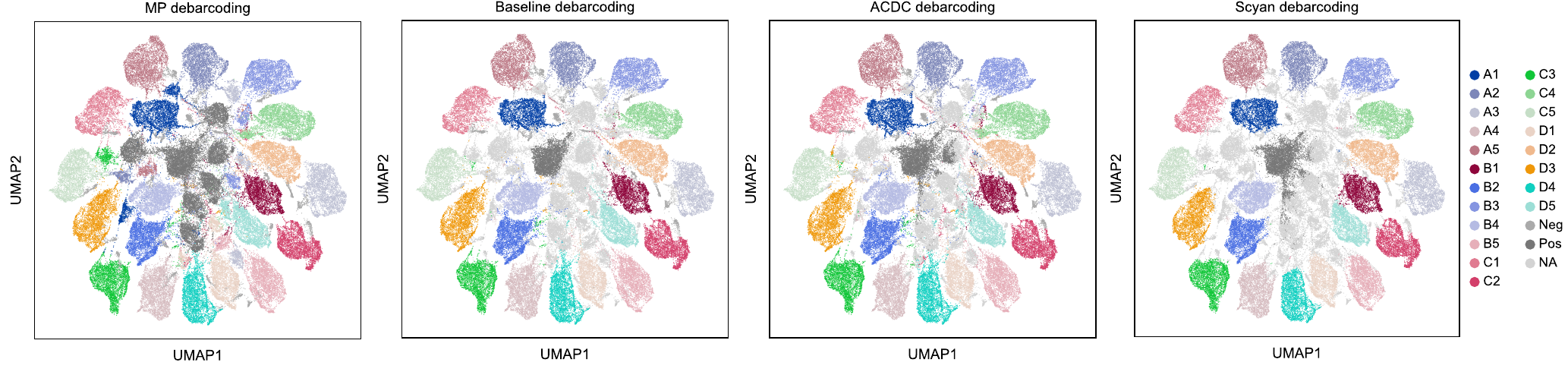}
\caption{\label{fig:sup_6}\textbf{UMAPs for the debarcoding task. From left to right: MP, Baseline, ACDC, Scyan.}}
\end{figure*}

\begin{figure*}[hbt!]
\centering
\includegraphics[width=\linewidth]{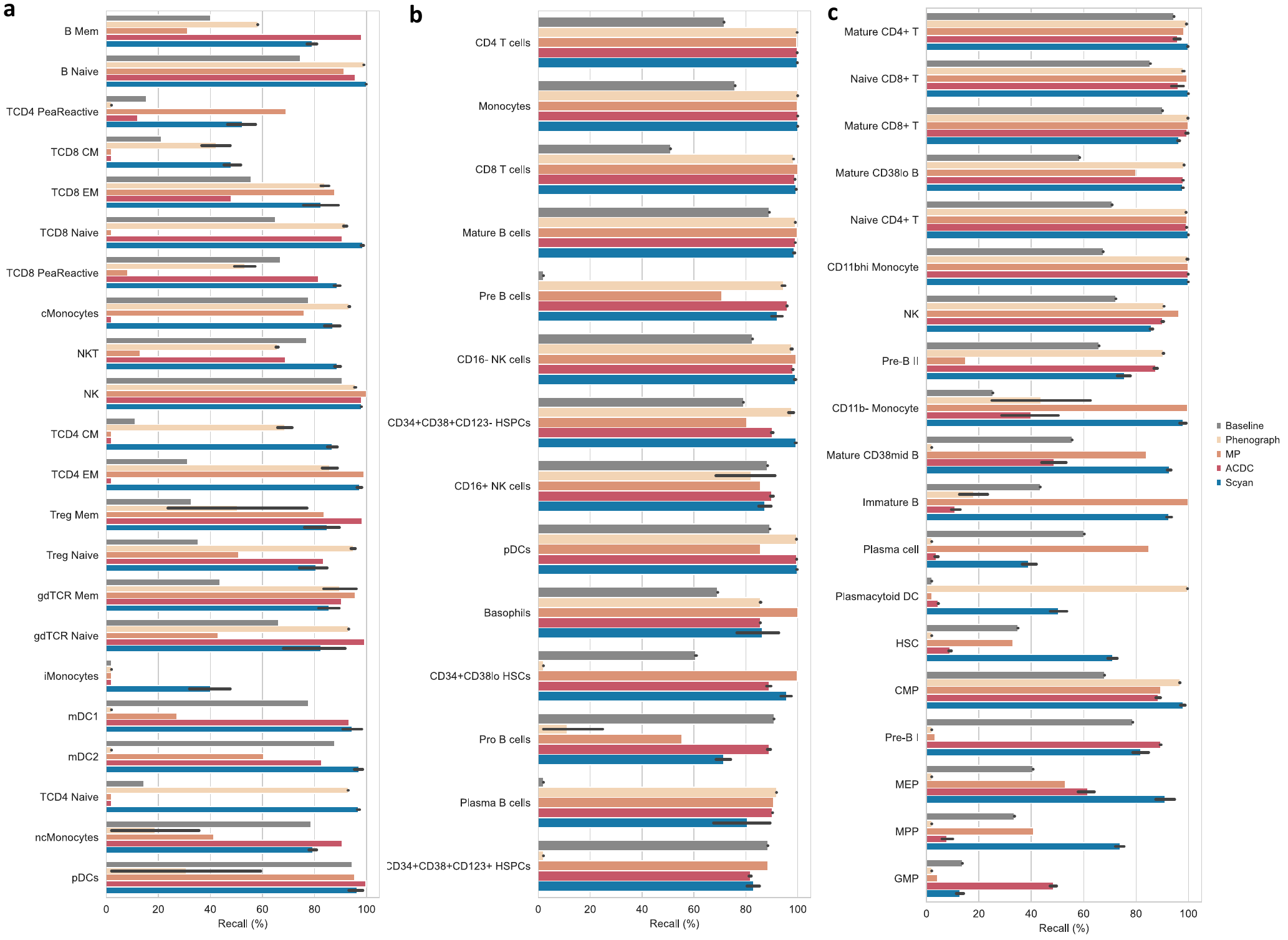}
\caption{\label{fig:sup_8}\textbf{Comparison of unsupervised models. The recall for each population is displayed} \textbf{a}, on POISED. \textbf{b}, on AML. \textbf{c}, on BMMC.}
\end{figure*}

\begin{figure*}[hbt!]
\centering
\includegraphics[width=\linewidth]{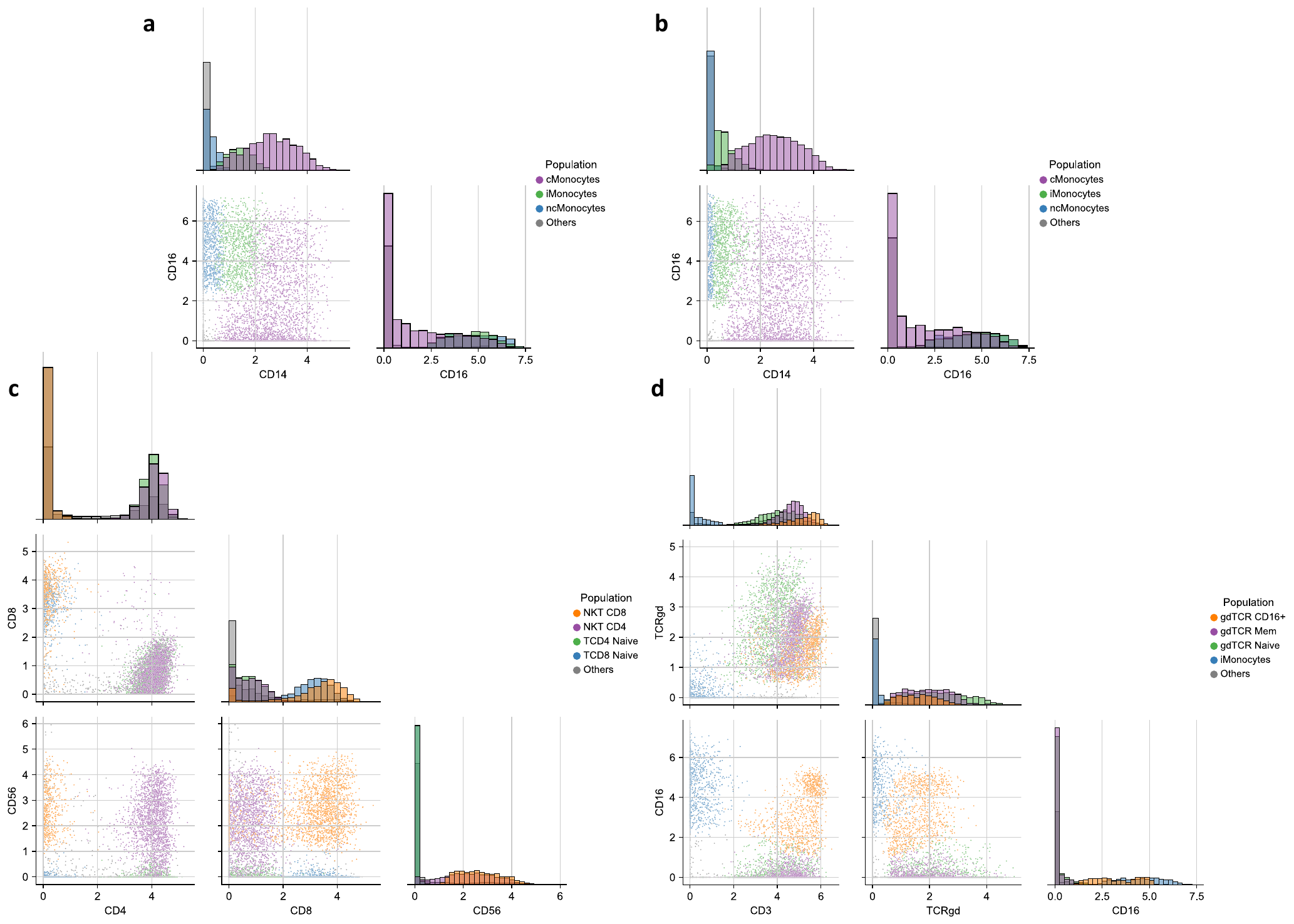}
\caption{\label{fig:sup_9}\textbf{Back-gating to check Scyan annotations} \textbf{a}, Scyan annotation among monocytes. \textbf{b}, Manual annotations among monocytes. \textbf{c}, NKT cells is a population that has been overpredicted by Scyan compared to manual gating. We check by back gating that they were properly annotated. \textbf{d}, gdTCR CD16+ is a new population discovered by Scyan. We show this population really exists.}
\end{figure*}

\begin{figure*}[hbt!]
\centering
\includegraphics[width=\linewidth]{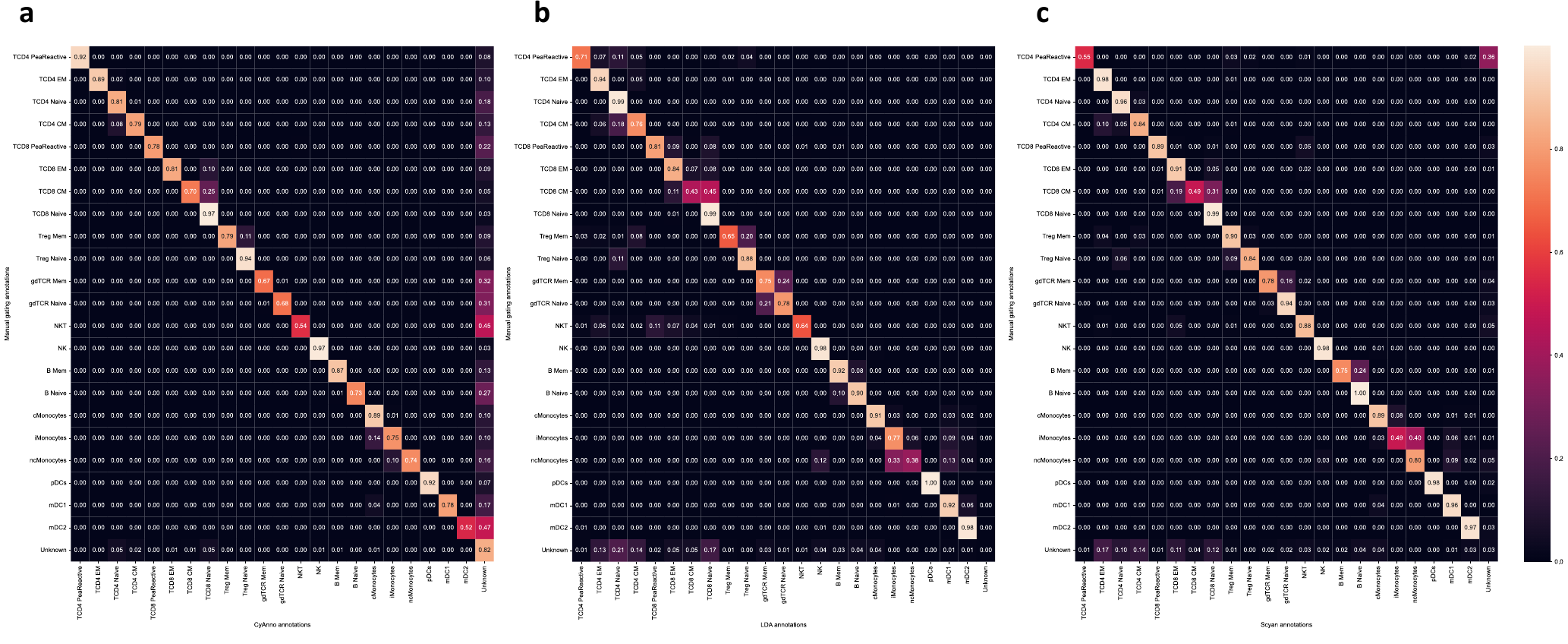}
\caption{\label{fig:sup_10}\textbf{Confusion matrix of annotations compared to manual gating.} \textbf{a}, CyAnno. \textbf{b}, LDA. \textbf{c}, Scyan.}
\end{figure*}

%% file: main.bbl
\section{References}
\begin{thebibliography}{1}
    \bibitem{cyto2} Behbehani, G. K. Immunophenotyping by Mass Cytometry. Methods Mol Biol 2032, 31–51 (2019).
\bibitem{cyto1} Spitzer, M. H. \& Nolan, G. P. Mass Cytometry: Single Cells, Many Features. Cell 165, 780–791 (2016).
\bibitem{flow_overview} McKinnon, K. M. Flow Cytometry: An Overview. Curr Protoc Immunol 120, 5.1.1-5.1.11 (2018).
\bibitem{cyto_limitations1} Newell, E. W. \& Cheng, Y. Mass cytometry: blessed with the curse of dimensionality. Nat Immunol 17, 890–895 (2016).
\bibitem{gating} Staats, J., Divekar, A., McCoy, J. \& Maecker, H. Guidelines for Gating Flow Cytometry Data for Immunological Assays. in Methods in molecular biology (Clifton, N.J.) vol. 2032 81–104 (2019).
\bibitem{nat_comment} Newell, E. W. \& Cheng, Y. Mass cytometry: blessed with the curse of dimensionality. Nat Immunol 17, 890–895 (2016).
\bibitem{critical_assessment} Aghaeepour, N. et al. Critical assessment of automated flow cytometry data analysis techniques. Nat Methods 10, 228–238 (2013).
\bibitem{phenograph} Levine, J. H. et al. Data-Driven Phenotypic Dissection of AML Reveals Progenitor-like Cells that Correlate with Prognosis. Cell 162, 184–197 (2015).
\bibitem{leiden} Traag, V. A., Waltman, L. \& van Eck, N. J. From Louvain to Leiden: guaranteeing well-connected communities. Sci Rep 9, 5233 (2019).
\bibitem{spade} Qiu, P. et al. Extracting a cellular hierarchy from high-dimensional cytometry data with SPADE. Nat Biotechnol 29, 886–891 (2011).
\bibitem{deepcytof} Li, H. et al. Gating mass cytometry data by deep learning. Bioinformatics 33, 3423–3430 (2017).
\bibitem{lda} Abdelaal, T. et al. Predicting Cell Populations in Single Cell Mass Cytometry Data. Cytometry Part A 95, 769–781 (2019).
\bibitem{cyanno} Kaushik, A. et al. CyAnno: a semi-automated approach for cell type annotation of mass cytometry datasets. Bioinformatics 37, 4164–4171 (2021).
\bibitem{review} Liu, P. et al. Recent Advances in Computer-Assisted Algorithms for Cell Subtype Identification of Cytometry Data. Front Cell Dev Biol 8, 234 (2020).
\bibitem{acdc} Lee, H.-C., Kosoy, R., Becker, C. E., Dudley, J. T. \& Kidd, B. A. Automated cell type discovery and classification through knowledge transfer. Bioinformatics 33, 1689–1695 (2017).
\bibitem{mp} Ji, D., Nalisnick, E., Qian, Y., Scheuermann, R. H. \& Smyth, P. Bayesian Trees for Automated Cytometry Data Analysis. in Proceedings of the 3rd Machine Learning for Healthcare Conference 465–483 (PMLR, 2018).
\bibitem{scvi} Lopez, R., Regier, J., Cole, M. B., Jordan, M. I. \& Yosef, N. Deep generative modeling for single-cell transcriptomics. Nat Methods 15, 1053–1058 (2018).
\bibitem{cellassign} Zhang, A. W. et al. Probabilistic cell-type assignment of single-cell RNA-seq for tumor microenvironment profiling. Nat Methods 16, 1007–1015 (2019).
\bibitem{saucie} Amodio, M. et al. Exploring single-cell data with deep multitasking neural networks. Nat Methods 16, 1139–1145 (2019).
\bibitem{nf1} Rezende, D. J. \& Mohamed, S. Variational Inference with Normalizing Flows. arXiv:1505.05770 [cs, stat] (2016).
\bibitem{nf2} Papamakarios, G., Nalisnick, E., Rezende, D. J., Mohamed, S. \& Lakshminarayanan, B. Normalizing Flows for Probabilistic Modeling and Inference. arXiv:1912.02762 [cs, stat] (2021).
\bibitem{nf3} Izmailov, P., Kirichenko, P., Finzi, M. \& Wilson, A. G. Semi-Supervised Learning with Normalizing Flows. arXiv:1912.13025 [cs, stat] (2019).
\bibitem{harmony} Korsunsky, I. et al. Fast, sensitive and accurate integration of single-cell data with Harmony. Nat Methods 16, 1289–1296 (2019).
\bibitem{cydar} Lun, A. T. L., Richard, A. C. \& Marioni, J. C. Testing for differential abundance in mass cytometry data. Nat Methods 14, 707–709 (2017).
\bibitem{combat} Johnson, W. E., Li, C. \& Rabinovic, A. Adjusting batch effects in microarray expression data using empirical Bayes methods. Biostatistics 8, 118–127 (2007).
\bibitem{bmmc} Bendall, S. C. et al. Single-Cell Mass Cytometry of Differential Immune and Drug Responses Across a Human Hematopoietic Continuum. Science 332, 687–696 (2011).
\bibitem{smote} Chawla, N. V., Bowyer, K. W., Hall, L. O. \& Kegelmeyer, W. P. SMOTE: Synthetic Minority Over-sampling Technique. jair 16, 321–357 (2002).
\bibitem{imblearn} Lemaître, G., Nogueira, F. \& Aridas, C. K. Imbalanced-learn: A Python Toolbox to Tackle the Curse of Imbalanced Datasets in Machine Learning. Journal of Machine Learning Research 18, 1–5 (2017).
\bibitem{umap} McInnes, L., Healy, J. \& Melville, J. UMAP: Uniform Manifold Approximation and Projection for Dimension Reduction. arXiv:1802.03426 [cs, stat] (2020).
\bibitem{tsne} van der Maaten, L. \& Hinton, G. Viualizing data using t-SNE. Journal of Machine Learning Research 9, 2579–2605 (2008).
\bibitem{debarcoding} Zunder, E. R. et al. Palladium-based mass tag cell barcoding with a doublet-filtering scheme and single-cell deconvolution algorithm. Nat Protoc 10, 316–333 (2015).
\bibitem{imc} Chang, Q. et al. Imaging Mass Cytometry. Cytometry A 91, 160–169 (2017).
\bibitem{logicle} Parks, D. R., Roederer, M. \& Moore, W. A. A new “Logicle” display method avoids deceptive effects of logarithmic scaling for low signals and compensated data. Cytometry Part A 69A, 541–551 (2006).
\bibitem{realnvp} Dinh, L., Sohl-Dickstein, J. \& Bengio, S. Density estimation using Real NVP. arXiv:1605.08803 [cs, stat] (2017).
\bibitem{temp1} Ackley, D. H., Hinton, G. E. \& Sejnowski, T. J. A learning algorithm for boltzmann machines. Cognitive Science 9, 147–169 (1985).
\bibitem{temp2} Ficler, J. \& Goldberg, Y. Controlling Linguistic Style Aspects in Neural Language Generation. http://arxiv.org/abs/1707.02633 (2017) doi:10.48550/arXiv.1707.02633.
\bibitem{mmd} Gretton, A., Borgwardt, K. M., Rasch, M. J., Schölkopf, B. \& Smola, A. A kernel two-sample test. J. Mach. Learn. Res. 13, 723–773 (2012).
\bibitem{adam} Kingma, D. P. \& Ba, J. Adam: A Method for Stochastic Optimization. http://arxiv.org/abs/1412.6980 (2017) doi:10.48550/arXiv.1412.6980.
\bibitem{dbs} Davies, D. \& Bouldin, D. A Cluster Separation Measure. Pattern Analysis and Machine Intelligence, IEEE Transactions on PAMI-1, 224–227 (1979).
\bibitem{silhouette} Rousseeuw, P. J. Silhouettes: A graphical aid to the interpretation and validation of cluster analysis. Journal of Computational and Applied Mathematics 20, 53–65 (1987).
\bibitem{pytorch} Paszke, A. et al. PyTorch: An Imperative Style, High-Performance Deep Learning Library. http://arxiv.org/abs/1912.01703 (2019) doi:10.48550/arXiv.1912.01703.
\bibitem{mediantrick} Garreau, D., Jitkrittum, W. \& Kanagawa, M. Large sample analysis of the median heuristic. http://arxiv.org/abs/1707.07269 (2018) doi:10.48550/arXiv.1707.07269.
\bibitem{sklearn} Pedregosa, F. et al. Scikit-learn: Machine Learning in Python. J. Mach. Learn. Res. 12, 2825–2830 (2011).
\bibitem{gan} Goodfellow, I. J. et al. Generative Adversarial Networks. (2014) doi:10.48550/arXiv.1406.2661.
\bibitem{vae} Kingma, D. P. \& Welling, M. An Introduction to Variational Autoencoders. FNT in Machine Learning 12, 307–392 (2019).
\bibitem{poised} Chinthrajah, R. S. et al. Sustained outcomes in oral immunotherapy for peanut allergy (POISED study): a large, randomised, double-blind, placebo-controlled, phase 2 study. The Lancet 394, 1437–1449 (2019).
\bibitem{tem} Sallusto, F., Lenig, D., Förster, R., Lipp, M. \& Lanzavecchia, A. Two subsets of memory T lymphocytes with distinct homing potentials and effector functions. Nature 401, 708–712 (1999).
  
\end{thebibliography}
